%% file: anomaly.tex
\documentstyle[preprint,aps,eqsecnum,floats,epsf]{revtex}

\begin{document}

\preprint {Submitted to Int. J. of Mod. Phys.}

\author{Lan Yin and Sudip Chakravarty}

\address{Department of Physics and Astronomy\\
University of California Los Angeles\\
Los Angeles, California 90095-1547}

\title{Spectral Anomaly and High Temperature Superconductors}
\date{November 16, 1995}

\maketitle

\tableofcontents
\newpage

\section*{}
\input{anomaly1}

\section{Introduction}
\input{anomaly2}

\section{Spectral anomaly}
\input{anomaly3}

\section{Perturbations}
\input{anomaly4}

\section{In-plane gap equation}
\input{anomaly5}

\section{Josephson effect}
\input{anomaly6}

\section{Interlayer pair tunneling}
\input{anomaly7}

\section{Spin-charge separation}
\input{anomaly8}

\section{Conclusions}
\input{anomaly9}

\section*{Acknowledgments}
We thank P. W. Anderson for numerous discussions. In addition, the chapters of
his forthcoming book, available to us for more than several years, have been
greatly useful. We also thank E. Abrahams for pointing out an error made by
one of us (S. C.) earlier.  A part of this work was carried out at the Aspen
center for physics. This work was supported by the National Science Foundation,
NSF-DMR-92-20416.
\appendix
\input{appendix}

\input {anomaly.BBL}
\end{document}

%% file: anomaly1.tex
\begin{abstract}

Spectral anomaly for interacting Fermions is characterized by the spectral function
$A([k-k_F],\omega)$ satisfying the scaling relation $A(\Lambda^{y_1}
[k-k_F],\Lambda^{y_2}\omega)= \Lambda^{y_A}A([k-k_F],\omega)$, where $y_1$,
$y_2$, and $y_A$ are the exponents defining the universality
class.  For a Fermi liquid $y_1=1$, $y_2=1$, $y_A=-1$; all other values
of the exponents are termed anomalous. In this paper, an   
example for which
$y_1=1$, $y_2=1$, but $y_A=\alpha-1$ is considered in detail. Attractive
interaction added to such a critical system leads to a novel superconducting
state, which is explored and its relevance to high
temperature cuprate superconductors is discussed.
\end{abstract} 
\pacs{}

%% file: anomaly2.tex
The normal state of  high temperature superconductors is unusual:
although the normal state has a metallic response and reflects  gaplessness, its
properties  cannot be explained by  Fermi
liquid theory\cite{Anderson1,Anderson2,Varma}. 
However, a  microscopic theory of such a gapless non-Fermi liquid in dimensions
$d\ge 2$ does not exist. In contrast, the breakdown of Fermi liquid
theory in $d=1$ has been known since the pioneering work of Mattis and
Lieb\cite{Mattis}. The absence of fermionic quasiparticles, spin-charge
separation, and power-law correlation functions have been extensively
discussed\cite{Emery,Schulz}, and Haldane\cite{Haldane} has
shown that these properties hold for a large class of interacting systems,  the
Luttinger liquids.

At the same time, a theory of high temperature superconductors has been
proposed\cite{Anderson,Chakravarty2} in which Josephson tunneling of the Cooper
pairs between the CuO  layers plays an important role, and leads, in particular,
to a striking dependence of the transition temperature on the number of such
layers in the unit cell and to a high transition temperature by exploiting, in a
very special way, the new energy scale proportional to the square of the tunneling
matrix element divided by the in-plane band width. The validity of this theory
depends on the non-Fermi liquid  nature of the normal state\cite{Chakravarty1}, as
the special momentum conserving Josephson pair tunneling Hamiltonian cannot
possibly be correct for a Fermi liquid. In view of the absence of a precise
microscopic theory of the non-Fermi liquid, it is necessary to explore the extent
to which a consistent theory can be developed based on general arguments because
the consequences of this interlayer tunneling theory are quite tantalizing. It is
our intention to present some general arguments in a reasonably complete form
and to point out directions in which further improvements are essential.

In the present paper, we shall {\sl assume}
the existence of a higher dimensional non-Fermi liquid state 
and study the  consequences for superconductivity. 
The essence of our arguments is a model of the
one-particle Green's function and the corresponding spectral function. 
In constructing this model,
we follow the stratagy of Galitskii and Migdal\cite{Migdal} who
constructed a model for a Fermi liquid and examined its self
consistency. The resulting theory is, of course, the same as the Landau
theory of Fermi liquids. In the Galitski-Migdal model, the  one-particle Green's
function, analytically continued to the  complex  plane, contains simple poles on a
second Riemann sheet, while the principal sheet is cut along the real frequency
axis. The quasiparticle is a resonance in a many-body state, resembling the  bare
particle.  In contrast, in our model, there are no poles but branch points on a
second sheet, located infinitesimally close to the real axis. Thus, the
non-Fermi liquid state is gapless and responds like a metal, but without
quasiparticles.
Such a model is  motivated by the Luttinger liquid concept,
but  the spectral function is only qualitatively similar to that of the Luttinger liquid.
In fact, in the present context, this model was briefly sketched by Wen\cite{Wen}.
The key element is the characterization of the non-Fermi
liquid state as a critical state defined by a scale invariant spectral
function\cite{Chakravarty1}. This characterization is described by the term spectral
anomaly, and, as in the theory of critical phenomena, the  universality classes are
distinguished by the critical exponents of the spectral function. 
Only a very special choice of the exponents leads to the fermi liquid model. 
In fact, we believe that
all systems that respond as metals can be be described by our model.
We also pay special attention to the Ward identity that relates the vertex
function to the inverse Green's function. This is essential in building a
consistent description of the non-Fermi liquid state.

The outline of our paper is as follows. In Sec. (II), we introduce the
concept of spectral anomaly and examine the nature of the normal state.
In Sec. (III), we consider the  effect of various interactions added to the
normal state described by  spectral anomaly. In this section, we discuss only in-plane
pairing interactions. In Sec. (IV), the superconducting gap equations with in-plane pairing
interactions are solved for both $s$-wave and $d$-wave kernels, to examine the
effect of the non-Fermi liquid behavior of the normal state.  Sec. (V) contains a
discussion of the Josephson effect and consititutes a qualitative justification
of the interlayer pair-tunneling Hamiltonian discussed in Sec. (VI). It is worth
noting that the analysis in this section bears some resemblance with a recent
paper by Khlebnikov\cite{Khlebnikov}. In Sec. (VI), we discuss the interlayer
tunneling model, and, compute tunneling density of states with parameters as
realistic as possible to compare with recent vacuum tunneling experiments. In
Sec. (VII) we consider spin-charge  separation and discuss how our basic model is
changed in its presence. In the final concluding section, Sec. (VIII), we point
out the directions in which our theory should be extended and improved.

%% file: anomaly3.tex
For a finite system, the K\"{a}llen-Lehmann representation\cite{Fetter} of the
time-ordered one-particle Green's function exhibits 
a discrete set of poles corresponding to the excitation frequencies of the
system. In the infinite volume limit, the poles merge into a cut that runs along the
real axis. Moreover, causality requires that the analytic continuation from the real
axis to the complex plane be free of singularities in the first and the third
quadrants. In fact, the only singularity on the principal
sheet is a cut along the real axis. 

To describe a Fermi liquid, Galitskii and
Migdal\cite{Migdal} proposed a model of the one-particle Green's function, which
contains singularities in the second and the fourth quadrants on a second sheet.
These singularities were chosen by them to be simple poles, sufficiently close to the
real axis to define long-lived quasiparticle states. The pole in the  fourth quadrant
corresponds to a quasiparticle and that in the second quadrant to a quasihole.
(Throughout the paper we choose to work with a fixed chemical potential $\mu$ in a
grand canonical ensemble rather than in a canonical ensemble for formal simplicity.).
This model corresponds to the assumption that
the matrix elements of the Fermion operators between the ground state and the exact
eigenstates are sharply peaked at an energy that depends on the
wavevector {\bf k}. The quasiparticle pole in the complex plane is the representation
of the physical phenomena on the real frequency axis, as the theory of analytic
continuation implies. 

A simple pole  corresponds to a unique energy-momentum
dispersion of the quasiparticle: with a wave vector $\bf k$, we associate an
energy $\varepsilon_{\bf k}$. And, with the renormalized wave function, the
quasiparticles can be made to resemble the bare particles for energies asymptotically
close to the Fermi energy. Other properties of the quasiparticles also follow from the
assumptions of the model.  
The charge of the quasiparticle is  unity and its spin  $1/2$.

The Galitskii-Migdal model can be summarized by the spectral function
\begin{eqnarray}
A(k,\omega)&=& -{1\over \pi}{\rm Im}G_{\rm R}(k,\omega) \nonumber \\
&=&z_k\delta(\omega-(\varepsilon(k)-\mu)),
\end{eqnarray}
where $G_{\rm R}$ is the retarded Green's function, and $z_k$ is the
quasiparticle residue. As this model is  valid for quasiparticles
close to the Fermi surface, we can linearize the spectrum and write $\varepsilon
(k)-\mu\approx v_F(k-k_F)$, where
$k_F$ is the Fermi wave vector and $v_F$ is the Fermi velocity. Then, the spectral
function has the scaling  property
\begin{equation}
A(\Lambda [k-k_F],\Lambda\omega)=\Lambda^{-1}A([k-k_F],\omega),
\end{equation}
reflecting the gaplessness of  the elementary
excitations. The system is poised at criticality. That a
normal Fermi system represents a critical system is, of course, understood.
The important point,
however, is that the criticality in a normal Fermi system does not usually lead to
serious dynamical consequences. This is  because the modes (the Fermionic excitations)
decouple, as in the Gaussian model of  classical critical phenomenon. 

We would like to build a model in which the criticality of the Fermi system is
more important. The simplest possible hypothesis is to assume that the analytic
continuation of the Green's function to a second sheet contains branch points instead
of simple poles. We postulate, therefore, a spectral function  that satisfies the
scaling relation
\begin{equation}
A(\Lambda^{y_1}[k-k_F],\Lambda^{y_2}\omega)= \Lambda^{y_A}A([k-k_F],\omega),
\end{equation}
where $y_1$, $y_2$, and $y_A$ are the  exponents defining the universality class of
the critical Fermi system. The values of the exponents other than the set
$y_1=1$, $y_2=1$, and $y_A=-1$ 
will be termed anomalous, while this special set  represents a Fermi liquid
for which the branch points collapse into simple poles.  The notion of a spectral
function with anomalous exponents will be termed {\sl spectral anomaly}. 

The anomalous 
spectral function  implies that a given $\bf k$ does not correspond to
a single frequency but a continuum of frequencies. The creation operator of wave vector
{\bf k}, applied to the ground state, creates a state that couples essentially to many
eigenstates that cannot be expressed by the model of a weakly damped quasiparticle.

To establish continuity with the Fermi liquid model, as the exponents approach
the Fermi liquid values, we make an important assumption. The location of the branch
point is assumed to be in one-to-one correspondence with the states of the
noninteracting system, which may be thought of as a remnant of Luttinger's theorem. This
is by no means an obvious assumption, but holds for one-dimensional electron gas
models\cite{Mattis}, and  also  for  gauge
models\cite{gauge} in higher dimensions.

From above, it is clear that we have not incorporated spin-charge separation in our
definition, for we have assumed that the spectral function is singular at a single
energy. The mathematical signature of spin-charge separation can be understood from the
properties of a one-dimensional Luttinger liquid where the electron
is a composite particle. The  elementary particles are spinons that
carry only spin and holons that carry only charge. As a result, the spectral function
is singular at two energies. A generalization to include spin-charge separation is 
discussed in Sec. VII.

From the dispersion relation
we can now determine the real part of the retarded Green's function, i.e,
\begin{equation}
{\rm Re}G_{\rm R}([k-k_F],\omega)=-{{\rm P}\over \pi}\int_{-\infty}^{+\infty}
d\omega'{{\rm Im}G_{\rm R}([k-k_F],\omega)\over \omega - \omega'}.
\end{equation}
Thus, the real part also satisfies the same
scaling relation given by
\begin{equation}
{\rm Re}G_{\rm R}(\Lambda^{y_1}[k-k_F],\Lambda^{y_2}\omega)=
\Lambda^{y_A}
{\rm Re}G_{\rm R}([k-k_F],\omega).
\end{equation}
From the generalized homogenity assumption, we also get
\begin{equation}
A([k-k_F],0)={1\over |k-k_F|^{y_A/y_1}}A([k-k_F]/|k-k_F|,0)
\end{equation}
and 
\begin{equation}
A(0,\omega)={1\over |\omega|^{y_A/y_2}}A(0,\omega/|\omega|).
\end{equation}
It is also possible to derive the momentum disrtibution function without any detailed
knowledge of the function $A(k,\omega)$. From the Lehmann representation,
the momentum distribution function $n(k)$ is given by
\begin{equation}
n(k-k_F)=\int_{-\infty}^0 d\omega A([k-k_F],\omega).
\end{equation}
Using the scaling relation, we can write
\begin{equation}
n(k-k_F)=\int_{-\infty}^0 d\omega
\Lambda^{-y_A}A(\Lambda^{y_1}[k-k_F],\Lambda^{y_2}\omega).
\end{equation}
Because $\Lambda$ is an arbitrary scale factor, we can chose
$\Lambda^{y_1}|k-k_F|=1$. Then, 
\begin{equation}
n(k-k_F)=|k-k_F|^{(y_2+y_A)/y_1}\int_{-\infty}^0 dx A({\rm Sign}(k-k_F),x).
\end{equation}
The  discontinuity present in a 
Fermi lquid  is destroyed due to spectral anomaly. In a Fermi system, the occupation
of a state, 
$n(k)$, cannot  diverge, and the
inequality 
$
(y_2+y_A)/ y_1 > 0 
$
must be satisfied. There is a superficial difficulty with the scaling argument above:
the integral for $n(k)$ does not converge. Thus, the  quantity
$n(k)$ cannot be obtained from the scaling argument, and we need to restore the cutoff.
However, we believe that the critical exponent is still given correctly by the
scaling argument. 

Consider now the time ordered Green's function, $G(k,\omega)$. It is
\begin{equation}
G(k,\omega)={\rm Re}G_R(k,\omega)-i{\rm Sign}(\omega){\rm Im}G_R(k,\omega).
\end{equation}
In terms of the time-ordered Greens function, we can write
\begin{equation}
n(k)={-i\over 2\pi}\lim_{t\to 0^+}\int_{-\infty}^{+\infty}d\omega
G(k,\omega)e^{i\omega t}.
\end{equation}
This integral can be broken up into three pieces. We write
\begin{equation}
n(k)={-i\over 2\pi}\lim_{t\to 0^+}\left[\int_{-\infty}^{-\omega_c}d\omega
G(k,\omega)e^{i\omega t}+\int_{-\omega_c}^{+\omega_c}d\omega
G(k,\omega)e^{i\omega t}+\int_{+\omega_c}^{+\infty}d\omega
G(k,\omega)e^{i\omega t}\right]
\end{equation}
The Green's function is assumed to be anomalous in the frequency range between
$-\omega_c$ and $\omega_c$. In the second integral, we can set
$t=0$  and apply the scaling form. The integral is well behaved
becuse the singularity,  as $\omega\to 0$, is softer than the Fermi liquid case
($\propto {1\over \omega}$).  The  first and the third
integrals  can be estimated by replacing  $G$ by its correct asymptotic
form, which is $G(k,\omega)\to {1\over
\omega}$, as
$\omega\to \infty$, resulting from the Fermion anticommutation
relation. Finally, we find again that
\begin{equation}
n(k)={1\over 2}-C\ {\rm Sign}(k-k_F)|k-k_F|^{(y_2+y_A)/y_1},
\end{equation}
where $C$ is a constant.

\subsection{The basic model}
To proceed further, we introduce an explicit model that is
defined by the retarded Green's function:
\begin{equation}
G_{\rm R}(k,\omega)={g(\alpha)e^{i\phi}\over
\omega_c^{\alpha}(\omega-\varepsilon (k)+i\delta)^{1-\alpha}}, \  -\omega_c \leq
\omega \leq \omega_c. 
\end{equation}
where $\phi$ is the phase, $\alpha$ is the 
anomaly exponent, and $\delta$ is an infinitesimal positive
number. To conform to causality, the upper half plane should be free
of singularities, and the cut is assumed to lie in the lower half
plane. The quantity $g(\alpha)$  will be determined from the
normalization. In a rough manner, one may include all nonsingular
self energy effects in the definition of $\varepsilon(k)$. Therefore,
\begin{equation}
-{1\over \pi}{\rm Im} G_{\rm R}(k,\omega)=-g(\alpha){\sin \phi\over
\pi}{\theta (\omega-\varepsilon (k))\over \omega_c^{\alpha}(\omega-\varepsilon
(k))^{1-\alpha}} -g(\alpha){\sin [\phi-\pi(1-\alpha)]\over
\pi}{\theta(\varepsilon (k)-\omega)\over
\omega_c^{\alpha}(\varepsilon(k)-\omega)^{1-\alpha}}.
\end{equation}
Because the spectral function has to be positive, the phase $\phi$ must
satisfy the inequality $0 > \phi >-\pi \alpha$. This condition is not
restrictive enough to fix $\phi$ uniquely. However, the time reversal symmetry is
violated unless $\phi=-\alpha \pi/2$. This is because of the invariance with
respect to $\omega \to -\omega$, and $\varepsilon (k)\to - \varepsilon (k)$.
Therefore, 
\begin{equation}
A(k,\omega)=g(\alpha){\sin ({\pi \alpha\over 2})\over \pi
\omega_c^{\alpha}}{1\over
|\varepsilon(k)-\omega|^{1-\alpha}}.\label{eq:spectrum}
\end{equation}

When $\varepsilon(k)=0$, our spectral function is similar to that of a
Luttinger liquid without spin-charge separation in $d=1$\cite{Meden,Voit}, but
it is very different when $\varepsilon(k)\ne 0$. In a Luttinger 
liquid, the spectral function diverges with a power law as
$\omega \to v_F k$ from above, vanishes for $-v_F k \leq \omega \leq  v_F k$, and 
vanishes with a power law as $\omega \to -v_F k$ from below.
In fact, the similarity with the 
one-dimensional system  cannot be expected. The special kinematic
constraints present in $d=1$ are missing  in higher dimensions. 
Finally, note also the difference with the gauge
models\cite{gauge}: although 
the spectral functions are similar for $\varepsilon(k)=0$, for
$k\ne 0$, the self energy in  the gauge models is nonsingular.

To preserve the equal-time anticommutation relation of the Fermions, one
must satisfy the sum rule $\int_{-\infty}^{\infty}d\omega A(k,\omega)=1$. If we
integrate the spectral function, given in Eq. (\ref{eq:spectrum}), we get
instead:
\begin{equation}
\int_{-\infty}^{\infty}d\omega A(k,\omega)=g(\alpha){\sin ({\pi \alpha\over 2})
\over \pi
\alpha}\left[\left(1+{\varepsilon(k)\over \omega_c}\right)^{\alpha}+
\left(1-{\varepsilon(k)\over \omega_c}\right)^{\alpha}\right].
\end{equation}
If we choose 
\begin{equation}
g(\alpha)= {\pi \alpha\over 2\sin ({\pi \alpha\over 2}) },
\end{equation}
the anticommutation relation will be
satisfied if $|\varepsilon(k)|\ll \omega_c$, which is precisely the regime in
which a scaling theory is appropriate.

We are still left with a minor problem with respect to the choice of the phase. For the
exact retarded Green's function,  the phase at $\omega=\infty$, $\phi
(\infty)=0-\varepsilon$, because the imaginary part   of $G_{\rm R}$ vanishes faster than
the real part. As $\omega$ runs along the real axis, ${\rm Im}G_{\rm R}$ does not
change sign and the phase remains in the lower half plane, reaching
$\phi(-\infty)=\pi+\varepsilon$. For our  Green's function,
the phase remains in the lower half plane, varying between $-\alpha\pi
/2$ and $(\pi +\alpha\pi/2)$. Clearly, the exact limits at
$\pm
\infty$ are violated, unless the scaling form is supplemented by the exact
asymptotic limits of the Green's function.  However, as long
as $\alpha \ne 0$,
discontinuities of the phase remain in our model; however, these disconitinuities
appear to be  physically inconsequential.

The time-ordered Green's function, $G(k,\omega)$, 
is given by 
\begin{equation}
G(k,\omega)={g(\alpha)\omega_c^{\alpha}}\left[{e^{-i{\pi
\alpha\over 2}}\theta(\omega)\over
(\omega-\varepsilon (k)+i\delta)^{1-\alpha}}+
{e^{i{\pi \alpha\over 2}}\theta(-\omega)\over
(\omega-\varepsilon (k)-i\delta)^{1-\alpha}}\right].
\end{equation}
Note that the argument of the $\theta$-function is the frequency variable,
$\omega$, and not $\varepsilon (k)$. For a Fermi liquid, there is no difference
between these two cases; this is not so here. The correct argument is
always $\omega$ (the total energy), and not the kinetic energy $\varepsilon (k)$.

\subsection{Momentum distribution function}
The momentum distribution function is now given by 
\begin{equation}
n(k)={\alpha\over2 \omega_c^{\alpha}}
\int_{-\omega_c}^0 d\omega {1\over |\varepsilon(k)-\omega|^{1-\alpha}}.
\end{equation}
We get
\begin{equation}
n(k)={1\over 2}\left[1-{\rm Sign}\left(\varepsilon(k)\right)
\left|{\varepsilon(k)\over \omega_c}\right|^{\alpha}\right].
\end{equation}
To understand what we meant previously by cutoff effects, let us attempt to
derive $n(k)$ by scaling and extending the limit to $-\infty$. Then,
\begin{equation}
n(k)={\alpha\over 2}
\left|{\varepsilon(k)\over \omega_c}\right|^{\alpha}\int_{-\infty}^0
dx {1\over |x-{\rm Sign}(\varepsilon(k))|^{1-\alpha}}.
\end{equation}
We see that the  singularity  of $n(k)$ is captured correctly, but not
the correct functional form. This is as it should be because the remaining
integral does not converge.
The calculation of $n(k)$ from the time-ordered Green's function
gives, as before, the same result.

\subsection{Matsubara Green's function}
The Matsubara Green's function, ${\cal G}(k,\omega_n)$, can be obtained from the
spectral function. Note that
\begin{equation}
{\cal G}(k,\omega_n)=\int_{-\infty}^{\infty}d\omega'{A(k,\omega')\over
i\omega_n-\omega'},
\end{equation}
where $\omega_n=\pi(2n+1) T$ are the Matsubara
frequencies; $T$ is the temperature. Substituting the spectral function, we get 
\begin{equation}
{\cal G}(k,\omega_n)=g(\alpha){\theta(\omega_n)e^{-i{\pi \alpha\over
2}}+
\theta(-\omega_n)e^{i{\pi \alpha\over 2}}\over
\omega_c^{\alpha}(i\omega_n-\varepsilon(k))^{1-\alpha}},
\end{equation}
where we have used the integral
\begin{equation}
\int_0^{\infty}dx {x^{\alpha-1}\over (z+x)}={\pi\over \sin (\pi\alpha)}
z^{\alpha-1},
\end{equation}
with $|{\rm Arg} (z)|<\pi$ and $1 >{\rm Re} \alpha > 0$.

\subsection{Density of states}
The density of states, $n(\omega)$, can be calculated from
the spectral function. It is 
\begin{equation}
n(\omega)={\alpha\over 2\omega_c^{\alpha}}
\int_{-\omega_c}^{\omega_c}d\omega'{\rho(\omega')\over
|\omega'-\omega|^{1-\alpha}}\label{eq:nofw},
\end{equation}
where
\begin{equation}
\rho(\omega)=\sum_k\delta(\varepsilon(k)-\omega).
\end{equation}
If we choose $\rho(\omega)\approx \rho=const.$ for $-W\le \omega \le W$ and 0
otherwise, then
\begin{equation}
n(\omega)={\rho\over 2}\left[\left({W\over \omega_c}+{\omega\over
\omega_c}\right)^{\alpha}+\left({W\over \omega_c}-{\omega\over
\omega_c}\right)^{\alpha}\right].
\end{equation}
We have also numerically calculated $n(\omega)$ for a more realistic
$\rho(\omega)$. For small $\alpha$, the smearing is not enough to destroy the van
Hove singularities, while for larger $\alpha$ it can completely smooth out the
singularities.

In contrast to Luttinger liquids, the density of states does
not vanish as $\omega \to 0$. In fact, it is essentially unchanged from the Fermi
liquid value, obtained as $\alpha\to 0$. It would agree exactly with the
Fermi liquid value if 
$\omega_c$ is chosen to be $W$.  Once
again, the correct treatment of the cutoff cannot be overemphasized. The scaling trick,
applied to the integral in Eq. (\ref{eq:nofw}) leads to
\begin{equation}
n(\omega)={\rho\alpha\over 2}\left|\omega\over \omega_c\right|^{\alpha}
\int_{-\infty}^{\infty}dx{1\over
|x-{\rm Sign}(\omega)|^{1-\alpha}},
\end{equation}
where the limits of the integral were set to $\pm \infty$. It would appear that
the density of states vanishes as $\omega\to 0$. This is incorrect,
because the remaining integral does not converge; the cutoffs cannot be set to
$\pm\infty$. The vanishing of the density of states at the Fermi energy in a
one-dimensional Luttinger liquid  is a
result of the special kinematic
constraints that do not easily generalize to higher dimensions.

\subsection{Consistency of the basic model}
It is useful to consider Fermi liquid theory as an
effective low energy theory of
metals\cite{Anderson,Benfatto,Shankar,Polchinski}. We shall briefly 
repeat the arguments due to Polchinski\cite{Polchinski}. Imagine that we
are considering an effective field theory below an energy scale
$\omega_c$ at which there are strong Coulomb interactions. We want to
find the effective low energy theory for energies smaller than $\omega_c$. The
elementary excitations of the effective field theory underlying the Galitski-Migdal
model are charged, spin-$1/2$, quasiparticles, similar to the original
electrons. 

Consider first the free action:
\begin{equation}
\int dt d^d{\bf p}\left\{ i\psi^{*}_{{\bf
p}\sigma}(t)\partial_t \psi_{{\bf
p}\sigma}(t)-\left(\varepsilon({\bf p})-\mu\right)\psi^{*}_{{\bf
p}\sigma}(t)\psi_{{\bf
p}\sigma}(t)\right\}
\end{equation}
As we scale all energies by a factor $s < 1$, the momenta must scale to
the Fermi surface. To accomplish this scaling, we write 
\begin{equation}
{\bf p}={\bf k}+{\bf l},
\end{equation}
where $\bf k$ is a vector on the Fermi surface and $\bf l$ is a vector
orthogonal to it. Let 
\begin{equation}
\varepsilon({\bf p}) - \mu = lv_F({\bf k})+O(l^2),
\end{equation}
where $v_F({\bf k})$ is the Fermi velocity. As we scale the energies and the wave
vectors,
$E\to sE$,
${\bf k} \to {\bf k}$, and
${\bf l}\to s{\bf l}$, the action remains fixed, provided  the
dimension of the fermion operators is
$s^{-1/2}$. Consider now all possible terms consistent with the
symmetries and examine their scaling. If there are no relevant operators, the
effective low energy theory is self-consistent. (The generated mass term 
can be absorbed by redefining the fermi surface to be  the true
interacting Fermi surface)

The most important interaction is the four fermion operator:
\begin{eqnarray}
\int dt&& d^{d-1}{\bf k_1}\ d{\bf l_1}d^{d-2}{\bf k_2}\ d{\bf l_2}
d^{d-1}{\bf k_3}\ d{\bf l_3} d^{d-1}{\bf k_4}\ d{\bf l_4} V({\bf k}_1,
{\bf k}_2,{\bf k}_3, {\bf k}_4)\\ \nonumber
&&\psi^{*}_{{\bf p}_1\sigma}(t)\psi_{{\bf p}_3\sigma}(t)
\psi^{*}_{{\bf p}_2\sigma'}(t)\psi_{{\bf p}_4\sigma'}(t)
\delta^d({\bf p}_1+{\bf p}_2-{\bf p}_3-{\bf p}_4),
\end{eqnarray}
where we have assumed that the interaction is short-ranged. It can be seen that
the interaction scales as
$s$ times the  dimension of the delta function.  Let us assume, for the moment, that
\begin{eqnarray}
\delta^d({\bf p}_1+{\bf p}_2-{\bf p}_3-{\bf p}_4)&=&
\delta^d({\bf k}_1+{\bf k}_2-{\bf k}_3-{\bf k}_4+{\bf l}_1+{\bf l}_2-{\bf
l}_3-{\bf l}_4) \nonumber \\
&\sim&\delta^d({\bf k}_1+{\bf k}_2-{\bf k}_3-{\bf k}_4),
\end{eqnarray}
where we have ignored $\bf l$ in comparison to $\bf k$. This justified because the
vectors $l$ scale to zero. Now the argument of the $\delta$-function does not depend on
$s$, and the four fermion interaction scales as $s$, vanishing in the limit $s\to
0$. The interaction is irrelevant. The conclusion is almost correct except for
the special kinematics implied by the Cooper channel
scattering processes. For these processes, the delta function can be seen to be of
order
$s^{-1}$, transforming the dimension of the interaction  to $s^0$. The net result
is that the interaction is marginal\cite{Polchinski}. 
Thus, except for special kinematics, the four
fermion interaction is irrelevant. The argument, valid in  dimensions $d\ge
2$, fails in $d=1$. In $d=1$, the $\delta$-function is always of order $s^{-1}$ and the
interaction is always marginal. 

We now turn to the consistency of our basic model. The action is
\begin{equation}
\int d\omega d^d{\bf p} G^{-1}({\bf p},\omega)\psi^{*}_{{\bf p}\sigma}(\omega)
\psi_{{\bf p}\sigma}(\omega),
\end{equation}
where the Green's function $G$ corresponds to that of the basic model. The dimension
of $\psi_{\sigma}({\bf p},\omega)$ is  $s^{-(3-\alpha)/2}$ and that
of  $\psi_{\sigma}({\bf p},t)$ is $s^{-(1-\alpha)/2}$. If we follow 
Polchinski, we see that the  the four fermion
interaction is irrelevant; even for those exceptional kinematics for which the
interaction was marginal in the Fermi liquid case, it now scales as $s^{2\alpha}$. The
spectral anomaly is more stable than the Fermi liquid! In fact, in the weak coupling
regime, it does not even allow a superconducting instability. In Sec. V, we shall see that this is indeed correct. The coupling has to reach a
threshold  before superconducting instability occurs. Remarkably enough, the Josephson
pair tunneling Hamiltonian\cite{Chakravarty2} describing interlayer tunneling is
relevant and results in a superconducting instability for arbitrarily weak
coupling. Such a term must therefore be included in the effective low energy
theory. 
\subsection{Vertex function}
In the normal state, the charge vertex and the current vertex satisfy the Ward identity\cite{Schrieffer}
\begin{equation}
\sum_\mu q_\mu \Gamma^\mu(k+\frac{q}{2},k-\frac{q}{2})
=G^{-1}(k+\frac{q}{2})-G^{-1}(k-\frac{q}{2}).
\end{equation}
In this subsection, we shall use the notation $q\equiv ({\bf q},q_0\equiv \omega)$ and the metric will be $(-1,1,1,1)$. Although this identity cannot fully determine
the vertex, it is a strong constraint.  In the limit $q \rightarrow 0$,
\begin{equation}
G^{-1}(k+\frac{q}{2})-G^{-1}(k-\frac{q}{2}) \approx \sum_\mu
	 q_\mu \partial_\mu G^{-1}(k) +O(q^3), 
\end{equation}
and the vertex can be chosen to be
\begin{equation}
\Gamma^\mu(k+\frac{q}{2},k-\frac{q}{2})= \partial_\mu G^{-1}(k)
\end{equation}
In our model, the $T=0$ Green's function is given by
\begin{equation}
G^{-1}(k)=g^{-1}(\alpha) \omega_c^\alpha e^{i\frac{\alpha\pi}{2}{\rm Sign}(k_0)}
(k_0-\varepsilon_{\bf k}+i\delta k_0)^{1-\alpha}
\end{equation}  
The vertex would be
\begin{eqnarray}
\Gamma^0(k+\frac{q}{2},k-\frac{q}{2})&=&
	\frac{(1-\alpha)e^{i\frac{\alpha\pi}{2}{\rm Sign}(k_0)} \omega_c^\alpha}
	{g(\alpha)(k_0-\varepsilon_{\bf k}+i\delta k_0)^\alpha}, \\
\Gamma^{i \not= 0} (k+\frac{q}{2},k-\frac{q}{2})&=&
	\frac{(1-\alpha)e^{i\frac{\alpha\pi}{2}{\rm Sign}(k_0)} \omega_c^\alpha}
	{g(\alpha)(k_0-\varepsilon_{\bf k}+i\delta k_0)^\alpha} v^i({\bf k})
\end{eqnarray}
However, the sign of the imaginary part of $G^{-1}(k)$ changes when $k_0=0$.
This leads to  a $\delta$-fuction contribution to $\Gamma^0$ but has no influence
on $\Gamma^{i \not= 0}$.  So, we have
\begin{eqnarray}
\lim_{q\rightarrow0} \Gamma^0(k+\frac{q}{2},k-\frac{q}{2})&=&
	\frac{\omega_c^\alpha} {g(\alpha)}
	\left[\frac{(1-\alpha)e^{i\frac{\alpha\pi}{2}{\rm Sign}(k_0)}}{(k_0-\varepsilon_{\bf k}+i\delta k_0)^\alpha}+
	2i\sin\frac{\alpha\pi}{2} |\varepsilon_{\bf k}|^{1-\alpha} \delta(k_0)\right], \\
\lim_{q\rightarrow0} \Gamma^{i \not= 0} (k+\frac{q}{2},k-\frac{q}{2})&=&
	\frac{(1-\alpha)e^{i\frac{\alpha\pi}{2}{\rm Sign}(k_0)} \omega_c^\alpha}
	{g(\alpha)(k_0-\varepsilon_{\bf k}+i\delta k_0)^\alpha} v^i({\bf k})
\end{eqnarray}
Note that the vertex functions are singular and the system is always strongly interacting.

%% file: anomaly4.tex
In the present section, we examine the effects of  perturbations. The word perturbation 
is meant in a specific sense. We augment the basic model by adding various 
interactions which we call
perturbations, but the solution of the resulting models is not perturbative. We consider two specific perturbations. The first is the interlayer tunneling in the normal state and the second is the in-plane BCS pairing interaction. The first perturbation is introduced to understand the incoherence of single particle tunneling. The second perturbation is designed to bring out certain features of the in-plane pairing. 

It is
convenient to use the language of functional integrals
The amplitude in imaginary time, $Z$,
is given by
\begin{equation}
Z\sim \int{\cal D}\psi_{\sigma}^*{\cal D}\psi_{\sigma}e^
{-S_0[\psi_{\sigma}^*,\psi_{\sigma}]-S_I[\psi_{\sigma}^*,\psi_{\sigma}]},
\end{equation}
where $S_0$ is
\begin{equation}
S_0=-{1\over \beta}\sum_{k,\sigma,\omega_n}
{\cal
G}^{-1}(k,\omega_n)\psi_{k\sigma}^*(\omega_n)\psi_{k\sigma}(\omega_n),
\end{equation}
where ${\cal G}(k,\omega_n)$ was derived earlier, and the spin index is denoted
by $\sigma$; $\beta$ is the inverse temperature.
$\psi^*$ and
$\psi$ are the Grassmann variables representing the fermions, and the Fourier series
\begin{equation}
\psi_{k\sigma}(\tau)={1\over
\beta}\sum_{\omega_n}e^{-i\omega_n\tau}\psi_{k\sigma}(\omega_n)
\end{equation}
relates the imaginary time Grassman variables to Grassman variables as a
function of Matsubara frequencies. 
The various perturbations are contained in
$S_I$.

\subsection{Interlayer tunneling in the normal state}
\label{sec-interlayer}
Consider two identical layers, $i=1,2$, that are coupled by a tunneling
matrix element, $t_{\perp}(k)$, assumed to be real. Then,
\begin{equation}
S_0=-{1\over \beta}\sum_{k,\sigma,\omega_n,i}
{\cal
G}^{-1}(k,\omega_n)\psi_{k\sigma}^{(i)*}(\omega_n)
\psi_{k\sigma}^{(i)}(\omega_n).
\end{equation}
The quantity $S_I$ is given by
\begin{equation}
S_I[\psi_{\sigma}^*,\psi_{\sigma}]={1\over \beta}\sum_{k,\sigma,\omega_n}
t_{\perp}(k)\left[\psi_{k\sigma}^{(1)*}(\omega_n)
\psi_{k\sigma}^{(2)}(\omega_n)+(1\to 2)\right].
\end{equation}
Now, $S_0+S_I$ can be expressed in a $2\times 2$ matrix form:
\begin{equation}
S_0+S_I=-{1\over
\beta}\sum_{k,\sigma,\omega_n}\bar{\psi}_{k\sigma}^{*}(\omega_n)
M(k,\omega_n)\bar{\psi}_{k\sigma}(\omega_n).
\end{equation}
The column vector $\bar{\psi}_{k\sigma}(\omega_n)$ is
\begin{equation}
\bar{\psi}_{k\sigma}(\omega_n)=\left(\begin{array}{c}
\psi_{k\sigma}^{(1)}(\omega_n)\\
\psi_{k\sigma}^{(2)}(\omega_n)\end{array}\right).
\end{equation}
The matrix $M(k,\omega_n)$ is
\begin{equation}
M(k,\omega_n)=\left(\begin{array}{cc}
{\cal G}^{-1}(k,\omega_n) & -t_{\perp}(k) \\
-t_{\perp}(k) & {\cal G}^{-1}(k,\omega_n)\end{array}\right).
\end{equation}

The  excitation spectrum is given by the inverse of the determinant, $[\det
M]^{-1}$, analytically continued to the real frequency axis. In particular,
the singularities of the time-ordered Green's function are given by
\begin{equation}
 [\det M]^{-1}={\theta(\omega)\over e^{i\pi\alpha} 
\Omega_+(k,\omega)^{2(1-\alpha)}-t_{\perp}^2(k)}+
{\theta(-\omega)\over e^{-i\pi\alpha} 
\Omega_-(k,\omega)^{2(1-\alpha)}-t_{\perp}^2(k)},
\end{equation}
where
\begin{equation}
\Omega_{\pm}(k,\omega)=[g^{-1}(\alpha)\omega_c^{\alpha}]^{1\over
(1-\alpha)} (\omega-\varepsilon(k)\pm i\delta) .
\end{equation}
There are branch points in the second and the fourth
quadrants of the complex plane at $\omega=\varepsilon(k)\pm i\delta$.
The cuts are shown in Fig. (1). In addition to the cuts there are poles. Let us consider the case $\omega > 0$.
\begin{figure}[htb]
\centerline{\epsfxsize 9cm
\epsffile{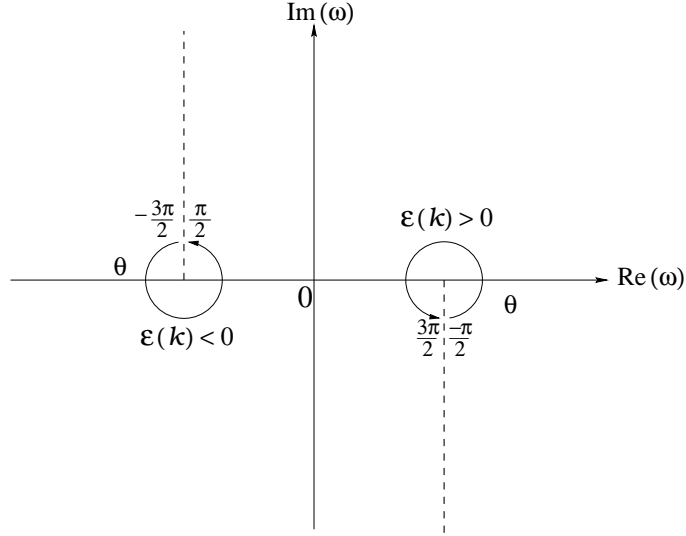}
}
\caption{Complex plane for the singularities of the Green's function for interlayer tunneling in the normal state.}
\label{f:1}
\end{figure}
The poles are given by
\begin{equation}
e^{i\pi\alpha} 
\Omega_+(k,\omega)^{2(1-\alpha)}=t_{\perp}^2(k)e^{2in\pi},
\end{equation}
where $n=0,\pm 1, \pm 2, \pm 3, \cdots$. If we define 
\begin{equation}
\omega-\varepsilon(k)+i\delta=|\omega-\varepsilon(k)|e^{i\theta},
\end{equation}
then the poles are the solutions of the following equations:
\begin{equation}
D^-(k,\omega)^{2(1-\alpha)}=t_{\perp}^2(k),
\end{equation}
\begin{equation}
\theta=-{\pi\alpha\over 2(1-\alpha)}+{n\pi\over (1-\alpha)},
\end{equation}
where
\begin{equation}
D^{\pm}(k,\omega)=[g^{-1}(\alpha)\omega_c^{\alpha}]^{1\over
(1-\alpha)} |\omega\pm\varepsilon(k)|
\end{equation}
To solve these equations, let us divide the complex plane into four regions
(I)-(IV) as shown in Fig. (2): $0 > \theta > -{\pi\over 2}$ (I), ${\pi\over 2} > \theta > 0$ (II), $\pi > \theta > {\pi\over 2}$ (III), and ${3\pi\over 2} > \theta > \pi$ (IV).
\begin{figure}[htb]
\centerline{\epsfxsize 9cm
\epsffile{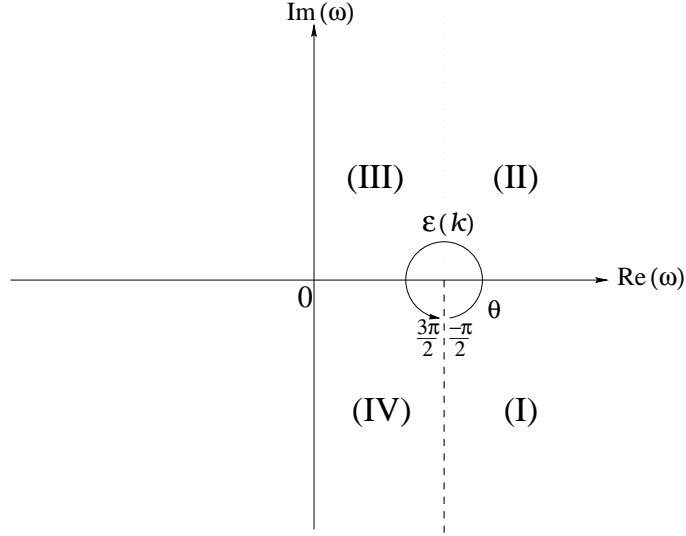}
}
\caption{Breakup into regions (I)-(IV) to solve for the poles of Green's function for interlayer tunneling in the normal state.}
\label{f:2}
\end{figure}
There are  no solutions in regions (II) and (III), as it
should be; otherwise, we would violate causality. The poles in regions (I) and (IV) are:
\begin{equation}
\omega^{\pm}_k=\varepsilon(k)+t_{\rm eff}(k)\left[\pm\sin{\pi/2\over 1-\alpha} + i
\cos{\pi/2\over 1-\alpha}\right],
\end{equation}
where
\begin{equation}
t_{\rm eff}(k)=g(\alpha)t_{\perp}(k)\left(g(\alpha){t_{\perp}(k)\over
\omega_c}\right)^{\alpha\over 1-\alpha}
\end{equation}
In the limit $\alpha\to 0$, these are the poles $\omega=\varepsilon(k)\pm
t_{\perp}(k)$. As $\alpha$ increases, the poles spiral in and become degenerate
at $\alpha=1/2$. Thus, at $\alpha=1/2$, no vestige of
coherence is left\cite{Strong}. For $\alpha > 1/2$, the poles fall off the Riemann
sheet. The poles for $\omega < 0$ are analogous.
Analytically
continued to real frequencies, the time ordered Green's function is,
\begin{eqnarray}
\left[G(k,\omega)\right]^{-1}=&&\theta(-\omega)\left(\begin{array}{cc}
e^{-i{\alpha\pi\over
2}}\Omega_-(k,\omega)^{1-\alpha} &
-t_{\perp}(k)
\\ -t_{\perp}(k) & e^{-i{\alpha\pi\over
2}}\Omega_-(k,\omega)^{1-\alpha}\end{array}\right)+
\nonumber \\&&\theta(\omega)\left(\begin{array}{cc}
e^{i{\alpha\pi\over
2}}\Omega_+(k,\omega)^{1-\alpha} &
-t_{\perp}(k)
\\ -t_{\perp}(k) & e^{i{\alpha\pi\over
2}}\Omega_+(k,\omega)^{1-\alpha}\end{array}\right).
\end{eqnarray}

\subsubsection{Spectral function}
The spectral function is 
\begin{equation}
A_{ij}(k,\omega)=-{1\over \pi} {\rm Sign}(\omega) {\rm Im}\left[G(k,\omega)\right]_{ij}
={1\over t_\perp(k)}F_{ij}\left(\frac{\omega-\varepsilon(k)}{t_{\rm eff}(k)}\right),
\end{equation}
where $F_{ij}(x)$ is a dimensionless function.
Its numerical evaluations  are shown in Figs. (3) and (4); the explicit expressions are
given in Appendix A. 
\begin{figure}[htb]
\centerline{\epsfxsize 9cm
\epsffile{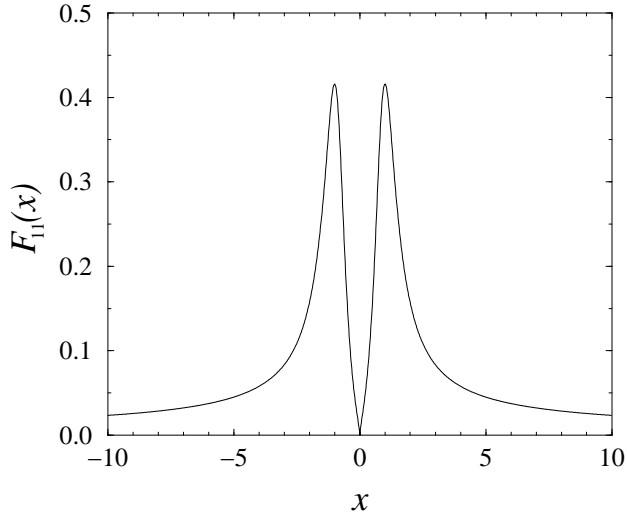}
}
\caption{Dimensionless diagonal spectral function for interlayer tunneling in the normal state;
$(\alpha=0.25)$.}
\label{f:3}
\end{figure}
 For the diagonal case, Fig. (3),  there  are two symmetrically
situated peaks, and the spectral weight vanishes at
$\varepsilon(k)$.
The shape of the peaks cannot be described by Lorentzians,
because, at high frequencies,  they are  easily seen to fall off as the power
law $\omega^{\alpha-1}$. Similarly, the off-diagonal spectral function is shown
in Fig. (4). The, shape, once again, is distinctly non-Lorentzian, falling off
as the power law
$\omega^{2\alpha-2}$.
\begin{figure}[htb]
\centerline{\epsfxsize 9cm
\epsffile{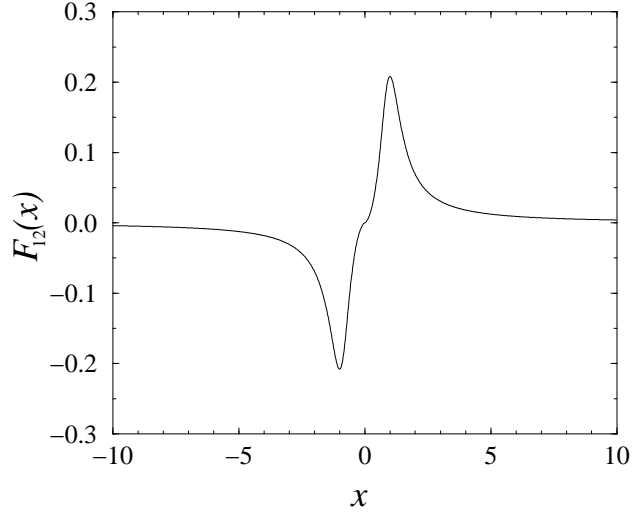}
}
\caption{Dimensionless off-diagonal spectral function for interlayer tunneling in the normal state.
$(\alpha=0.25)$}
\label{f:4}
\end{figure}
We have also calulated the corresponding density of
states. For the diagonal part, it is almost constant over the entire region
between $\omega_c$ and $-\omega_c$. For the off-diagonal part, it is essentially
zero over the same range. 

\subsection{In-plane pairing interactions}
Consider now a two-dimensional layer in which the electrons interact through an
attractive pairing interaction. Later, we shall generalize to include bilayer
systems. The interaction part,
$S_I[\psi_{\sigma}^*,\psi_{\sigma}]$, is
\begin{equation}
S_I=-{1\over
\beta^3}\sum_{{\bf k},{\bf k'},\omega,\omega_1,\omega_2}
V_{{\bf k},{\bf k'}}\psi_{{\bf k}\uparrow}^{*}(\omega_1)
\psi_{-{\bf k}\downarrow}^{*}(\omega-\omega_1)
\psi_{-{\bf k'}\downarrow}(\omega-\omega_2)
\psi_{{\bf k'}\uparrow}(\omega_2)
\end{equation}
For simplicity, the pairing interaction, $V_{k,k'}$, will be assumed to be
separable, i.e. ,
\begin{equation}
V_{k,k'}=\lambda_k \lambda_{k'}.
\end{equation}
Different pairing symmetries can be incoroprated by different choices of the
functions $\lambda_k$. Introducing an auxiliary complex field,
$c(\omega)$, and carrying out a Hubbard-Stratonovich transformation\cite{Popov},
we can write
\begin{equation}
Z\sim \int{\cal D}c {\cal D}c^*{\cal D}\psi_{\sigma}^*{\cal D}\psi_{\sigma}e^
{-\sum_{\omega}c^*(\omega)c(\omega)-S_0[\psi_{\sigma}^*,\psi_{\sigma}]-S_I[\psi_{\sigma}^*,\psi_{\sigma}]}
\end{equation}
The Grassmann variables can be integrated out by a linear shift, and we get
\begin{equation}
Z\sim \int{\cal D}c {\cal D}c^{*}e^{-\sum_{\omega}c^{*}(\omega)c(\omega)
+{\rm Tr}\ln A[c^{*}(\omega),c(\omega)]},
\end{equation}
where the matrix $A$ is
\begin{equation}
A^{\omega_1,\omega_2}_{k_1,k_2}=\delta_{{\bf k}_1,{\bf
k}_2}\left(\begin{array}{cc} {\cal
G}^{-1}({\bf k}_1,\omega_1)\delta_{\omega_1,\omega_2} & {\lambda_{k_1}\over
\sqrt{\beta}}c(\omega_1-\omega_2)\\
{\lambda_{k_1}\over
\sqrt{\beta}}c^*(\omega_2-\omega_1)
 & -{\cal G}^{-1}(-{\bf
k}_1,-\omega_1)\delta_{\omega_1,\omega_2}\end{array}\right)
\end{equation}
The saddle point corresponds to $c(\omega)=0$, $c^*(\omega)=0$, for $\omega\ne
0$. Then, denoting  the superconducting gap, $\Delta_k$,
by
\begin{equation}
\Delta_k={\lambda_{k}\over\sqrt{\beta}}c(0),
\end{equation}
we get
\begin{equation}
\Delta_k=\lambda_k\sum_{k'}\lambda_{k'}\Delta_{k'}\chi_p(k'),
\end{equation}
where the pair susceptibility, $\chi_p(k)$, is
\begin{equation}
\chi_p(k)=-{1\over \beta}\sum_{\omega_n}{1\over {\cal G}^{-1}({\bf k},\omega_n)
{\cal G}^{-1}(-{\bf k},-\omega_n)+\Delta_k^2}.
\end{equation}

The excitation spectrum is determined by the singularities of the analytic
continuation of $[\det A]^{-1}$, which is
\begin{equation}
[\det A]^{-1}=-{1\over
(\varepsilon^2(k)-\omega^2-i\delta)^{1-\alpha}\omega_c^{2\alpha}g^{-2}(\alpha)+\Delta_k^2}.
\end{equation}
In an earlier publication\cite{Chakravarty1} the phase was chosen
incorrectly; the mistake was pointed out to us by Abrahams\cite{Abrahams}.
Unfortunately, the same  error appears to have propagated to later
publications\cite{Sudbo,Muthukumar}. 

Clearly, there are branch points at
$\omega=\pm\varepsilon(k)$. In addition, the spectrum consists of quasiparticle
poles in the superconducting state. Consider 
$\varepsilon(k) > 0$, and divide the complex plane into six regions as shown in
Fig. (5). 
\begin{figure}[htb]
\centerline{\epsfxsize 9cm
\epsffile{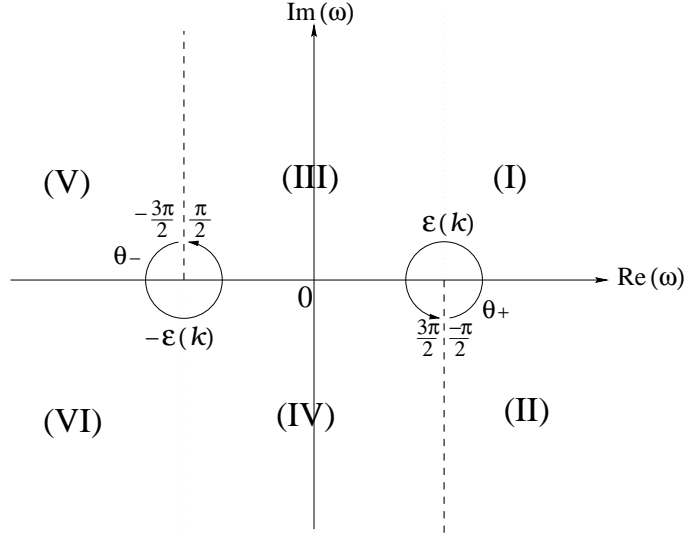}
}
\caption{Breakup of the complex plane into regions (I)-(VI) to obtain the poles of the superconducting Green's function.}
\label{f:5}
\end{figure}
Let,
\begin{equation}
(\omega-\varepsilon(k)+i\delta)^{1-\alpha}=|\omega-\varepsilon(k)|^{1-\alpha}
e^{i(1-\alpha)\theta_+}\ ,\  -{\pi\over 2} < \theta_+ < {3\pi\over 2}\ ,
\end{equation}
and,
\begin{equation}
(\omega+\varepsilon(k)-i\delta)^{1-\alpha}=|\omega+\varepsilon(k)|^{1-\alpha}
e^{i(1-\alpha)\theta_-}\ ,\  -{3\pi\over 2} < \theta_- < {\pi\over 2}\ .
\end{equation}
The quasiparticle poles are given by the solutions of the following equations:
\begin{equation}
|\omega^2-\varepsilon^2(k)|^{1-\alpha}\omega_c^{2\alpha}g^{-2}(\alpha)=\Delta_k^2
,
\end{equation}
and
\begin{equation}
\theta_++\theta_-=-{\alpha \pi\over 1-\alpha}+{2n\pi\over 1-\alpha},
\end{equation}
where $n=0,\pm 1, \pm 2, \pm 3, \cdots$
There are solutions only in regions (II) and (V). These are given by
\begin{equation}
\omega^2 =\varepsilon^2(k)+\Delta_{\rm eff}^2(k)e^{-i{\alpha\pi\over 1-\alpha}},
\end{equation}
where
\begin{equation}
\Delta_{\rm eff}(k)=g(\alpha)\Delta_k\left(g(\alpha){\Delta_k\over
\omega_c}\right)^{\alpha\over 1-\alpha}.
\end{equation}
The quasiparticles are both damped and shifted in energy due to  spectral
anomaly. The effective gap $\Delta_{\rm eff}(k)$ collapses at $\alpha=1$.

Analytically continued to real frequencies, the time ordered superconducting Greeen's
function at $T=0$ is 
\begin{eqnarray}
\left[G_s({\bf
k},\omega)\right]^{-1}=&&\theta(-\omega)\left(\begin{array}{cc}
e^{-i{\alpha\pi\over 2}}\Omega_-({\bf k},\omega)^{1-\alpha} &
\Delta_{\bf k}
\\ \Delta_{\bf k} & -e^{i{\alpha\pi\over
2}}\Omega_+(-{\bf
k},-\omega)^{1-\alpha}\end{array}\right)+
\nonumber \\&&\theta(\omega)\left(\begin{array}{cc}
e^{i{\alpha\pi\over
2}}\Omega_+({\bf k},\omega)^{1-\alpha} &
\Delta_{\bf k}
\\ \Delta_{\bf k} & -e^{-i{\alpha\pi\over
2}}\Omega_-(-{\bf
k},-\omega)^{1-\alpha}\end{array}\right).
\end{eqnarray}
\subsubsection{Superconducting spectral function}
The superconducting spectral function is found to be
\begin{equation}
A_s(k,\omega)={1\over |\Delta_k|}F_s(\frac{\omega}{|\Delta_{\rm eff}(k)|},\frac{\varepsilon(k)}{|\Delta_{\rm eff}(k)|}),
\end{equation}
where $F_s(x,y)$ is dimensionless.
The explicit expression is given in Appendix
B. In Fig. (6) we show the dimensionless spectral function for
$\varepsilon(k) > 0$; for 
$\varepsilon(k) < 0$, the figure should be symmetrically reflected. When
$\varepsilon(k) = 0$, the spectral function has two symmetrical peaks, in
contrast to two  delta functions  at $\pm\Delta_k$, as in BCS theory.
\begin{figure}[htb]
\centerline{\epsfxsize 9cm
\epsffile{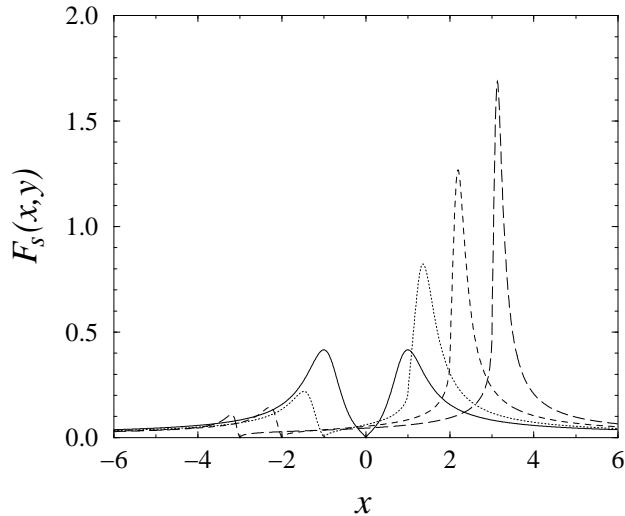}
}
\caption{Dimensionless superconducting spectral function.  The solid, dotted, dashed, long-dashed lines are for $y=0, 1, 2, 3$; $(\alpha=0.25)$.}
\label{f:6}
\end{figure}
As the energy increases, the weight of the peak on the positive frequency side
grows and that on the negative frequency side decreases. The weight vanishes
exactly at $-\varepsilon(k)$, but is finite everywhere else in contrast to BCS
theory. The location of the peaks are the real parts of the quasiparticle
energy.  The energy dependence cannot be described by Lorentzians
due to inherent power laws of the model.

%% file: anomaly5.tex
In this section we shall solve the gap equation that was derived in the previous section. We want to demonstrate that a critical value of the coupling is necessary if we assume only a in-plane pairing mechanism, and therefore by itself it is not such a promising mechanism for superconductivity if the normal state is a non-Fermi liquid. We also want to show that the density of states at zero frequency is finite even though the mean field theory leads to a pair amplitude. This is because the density of states is proportional to the momentum integral of the diagonal spectral function and the spectrum still contains gapless excitations arising from the cut in the Green's function. In solving the gap equations, it did not appear important to use
realistic one-electron band structures and the approximation of a circular Fermi surface is used.
\subsection{S-Wave}
Consider a pairing interaction that is attractive around the Fermi surface
over an energy range $\omega_D$ smaller than the cutoff $\omega_c$; for s-wave,
\begin{equation}
V_{k,k'}=-\lambda.
\end{equation} 
Then, the $T=0$ gap equation is
\begin{equation}
\Delta_0={\rho \lambda\over 2\pi}\int_{-\infty}^{\infty} d\varepsilon
\int_{-\omega_D}^{\omega_D}dy {\Delta_0\over
(y^2+\varepsilon^2)^{1-\alpha}\omega_c^{2\alpha}g^{-2}(\alpha)+\Delta_0^2}.
\end{equation}
For
simplicity, we have chosen a cutoff prescription in which the integration over
$\varepsilon$ is allowed to run between $\pm \infty$, but the remaining
frequency integration is restricted between $\pm \omega_D$.  Then, for $\alpha
< 1/2$, when $\Lambda\to \infty$, the gap
$\Delta_0$ is  found to be
\begin{equation}
\Delta_0=\omega_cg^{-1}(\alpha)\left(\omega_D\over
\omega_c\right)^{1-\alpha}\left[{1\over \pi}B\left({1\over 2},{1\over
2}-\alpha\right)\right]^{{1-\alpha\over 2\alpha}}\left[{(1-\alpha)\over
\pi\alpha}\sin\left({\pi\alpha\over1-\alpha}\right)\left(1-{\lambda_c\over
\lambda}\right)\right]^{1-\alpha\over 2\alpha},
\end{equation}
where 
\begin{equation}
\rho\lambda_c={2\alpha\pi g^{-2}(\alpha)\over B\left({1\over 2},{1\over
2}-\alpha\right)}\left({\omega_c\over\omega_D}\right)^{2\alpha}
\end{equation}
is the critical value of the coupling necessary to obtain a non-zero
superconducting gap, and $B(x,y)$ is the Eulerian beta function. Note that this critical value vanishes as $\alpha\to 0$,
and one recovers the BCS result
\begin{equation}
\lim_{\alpha\to 0} \Delta_0 = 2\omega_De^{-{1\over\rho\lambda}}.
\end{equation}

Consider now the equation for the transition temperature, $T_c$:
\begin{equation}
1={\lambda\over \beta_c}\sum_{\omega_n,k}{1\over
\left[\omega_n^2+\varepsilon^2(k)\right]^{1-\alpha}\omega_c^{2\alpha}g^{-2}(\alpha)}.
\end{equation}
Once again,  we choose to introduce cutoffs
on the Matsubara frequencies at $\pm \omega_D$, but let the integration over
$\varepsilon$ run over $\pm\infty$. Thus,
\begin{equation}
1={\rho\lambda\over \beta_c}\int_{-\infty}^{\infty}d\varepsilon\sum_n{1\over
\omega_c^{2\alpha}g^{-2}(\alpha)\left(\omega_n^2+\varepsilon^2\right)^{1-\alpha}}.
\end{equation}
We find that the 
that the transition temperature, $T_c$, is 
\begin{equation}
T_c={\omega_D\over \pi}\left({1\over 2^{2\alpha}-4\alpha}\right)^{1\over 2\alpha}
\left(1-{\lambda_c\over \lambda}\right)^{1\over 2\alpha},
\end{equation}
where the critical value of $\lambda$ is given by
\begin{equation}
\rho\lambda_c=\left(\omega_c\over
\omega_D\right)^{2\alpha}{2\alpha\pi g^{-2}(\alpha)\over B({1\over 2}, {1\over
2}-\alpha)}.
\end{equation}
The existence of the critical value of $\lambda$ was first pointed out by
Balatsy\cite{Balatsky}, although his discussion of the reentrant phase
transition appears to be incorrect due to an incorrect treatment of the cutoff.
In the limit $\alpha \to 0$, $\lambda_c \to 0$, and
\begin{equation}
T_c={e^2\omega_D\over 2\pi}e^{-{1\over \rho\lambda}}.
\end{equation}
The $4\%$ discrepancy in the prefactor in comparison to the BCS is result is
due to the different choice of the cutoff. This result disagrees with those
obtained by Sudb{\o}\cite{Sudbo} and Muthukumar {\em et al.\/}\cite{Muthukumar}. Our
calculation yields  the BCS result in the limit $\alpha\to 0$.
 
\subsubsection{Density of states for $s$-wave}
In Fig. (7) we show the superconducting density of states for various values
of the parameters; see Appendix B for details. The surprising feature is the gaplessness, although a
pseudogap persists. The square-root singularity of BCS is gone, but the peak is
located at the effective gap. 
\begin{figure}[htb]
\centerline{\epsfxsize 9cm
\epsffile{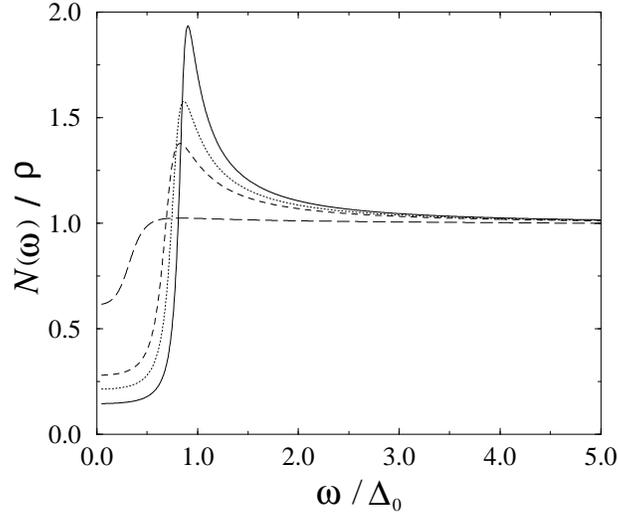}
}
\caption{Superconducting $s$-wave density of states for in-plane pairing.
The solid, dotted, dashed, long-dashed curves are for $\alpha=0.05, 0.075, 0.01, 0.25$.}
\label{f:7}
\end{figure}
The low frequency behavior of the density of states $N(\omega)$ for
$\alpha < 1/2$ is 
\begin{eqnarray}
\frac{N(\omega)-N(0)}{\rho}=&&\left[{\alpha(2^{1-\alpha}-1)\over
2-\alpha}\left({\omega_c\over
g(\alpha)\Delta_0}\right)^{\alpha}\right]\left({\omega\over
g(\alpha)\Delta_0}\right)^{2-\alpha}
\nonumber \\
&+&\left[{\alpha\over 2}\left(g(\alpha){\Delta_0\over
\omega_c}\right)^{\alpha\over 1-\alpha}B\left({2-3\alpha\over 2(1-\alpha)},
{2-\alpha\over 2(1-\alpha)}\right)\right]\left({\omega\over
g(\alpha)\Delta_0}\right)^2,
\end{eqnarray}
where
\begin{equation}
{N(0)\over \rho}=1-\frac{\left(\pi\alpha/2(1-\alpha)\right)}
{\sin\left(\pi\alpha/2(1-\alpha)\right)}\left(g(\alpha){\Delta_0\over
\omega_c}\right)^{\alpha\over 1-\alpha}.
\end{equation}
The frequency dependence of the first term is nonanalytic and the
coeffecient is essentially proportional to $\alpha$ for small $\alpha$;
the cutoff  and the gap dependences are weak. For, small $\alpha$, the quadratic
term competes with the first term. The background, $N(0)$, vanishes in the limit $\alpha\to 0$, and one recovers
the BCS result. The background also increases as $\alpha$ increases. 
Superconductivity  is of course preserved because the saddle point of the
action corresponds to broken gauge symmetry.

\subsection{D-Wave}
The zero temperature d-wave gap equation can be easily solved if we assume a
separable kernel, where
\begin{equation}
V_{k,k'}=-\lambda \cos2\theta_k\cos2\theta_{k'}
\end{equation} 
and $\lambda > 0$. Then the gap $\Delta_k$ is given by $\Delta_k=\Delta_0\cos
2\theta_k$, and $\Delta_0$, in turn, is 
\begin{equation}
\Delta_0=\omega_cg^{-1}(\alpha)\left(\omega_D\over
\omega_c\right)^{1-\alpha}\left[{(1-\alpha)B\left({1\over2},{1\over
2}-\alpha\right)\over 2\alpha\pi B
\left({1\over
2},{3-\alpha\over
2(1-\alpha)}\right)}\sin\left({\pi\alpha\over1-\alpha}\right)\left(1-{\lambda_c\over
\lambda}\right)\right]^{1-\alpha\over 2\alpha},
\end{equation}
where
\begin{equation}
\rho\lambda_c=4\alpha\left({\omega_c\over
\omega_D}\right)^{2\alpha}{\pi g^{-2}(\alpha)\over B\left({1\over2},{1\over
2}-\alpha\right)}.
\end{equation}

The superconducting transition temperature can be found as before and it is 
\begin{equation}
T_c={\omega_D\over \pi}\left[{1\over 2^{2\alpha}-4\alpha}\left(1-{\lambda_c\over
\lambda}\right)\right]^{1\over 2\alpha}
\end{equation}
The critical coupling, $\lambda_c$, is twice the s-wave critical temperature.
 
\subsubsection{Density of states for $d$-wave}
The superconducting density of states $N(\omega)$ is shown in Fig. (8). The logarithmic
peak at the amplitude of the gap is smeared out and there are states at zero
energy. The effect of spectral anomaly increases as $\alpha$ increases.
Consider now the low frequency behavior.
\begin{figure}[htb]
\centerline{\epsfxsize 9cm
\epsffile{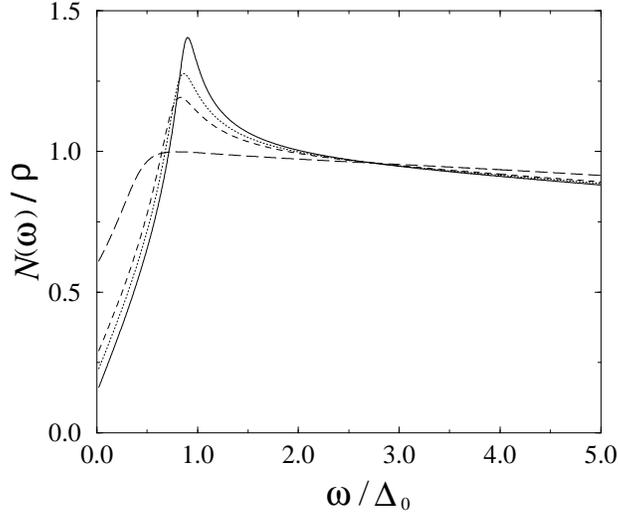}
}
\caption{Superconducting $d$-wave density of states for in-plane pairing.
The solid, dotted, dashed, long-dashed curves are for $\alpha=0.05, 0.075, 0.01, 0.25$.}
\label{f:8}
\end{figure}
For low frequencies, the integrals can be approximated and the result is
\begin{eqnarray}
{N(\omega)-N(0)\over\rho}&\approx&\left[{\alpha\over \sin({\alpha\pi\over 2})}
B\left({3-\alpha\over 2},{1+\alpha\over 2}\right)-{\alpha(1+2\alpha-\alpha^2)\over
2}\right]\left({\omega\over g(\alpha)\Delta_0}\right)\nonumber \\
&\approx&\left[1-{3\alpha\over 2}\right]\left({\omega\over g(\alpha)\Delta_0}\right).
\end{eqnarray}
Unlike s-wave, the low frequency exponent is unaffected by $\alpha$. The background
term is 
\begin{equation}
N(0)=\rho\left[1-{1\over 2}\left(g(\alpha){\Delta_0\over \omega_c}\right)^{\alpha\over
1-\alpha}B\left({1\over 2(1-\alpha)},1-{\alpha\over 2(1-\alpha)}\right)\right].
\end{equation}

\subsection{Collective mode}
In the superconducting state, the pairing interaction not only gives rise to
the gap, but also renormalizes the vertex and produces collective modes;
possible long range interactions are important as well.
In this section, we consider only the pairing interaction 
to examine the effect of non-fermi-liquid correlations.

In the random phase approximation and in the finite-temperature formalism, the vertex equation in the
$T\rightarrow0$ limit is given by\cite{Schrieffer}
\begin{equation}
{\hat \Gamma}_s(k,k+q)={\hat \Gamma}_n(k,k+q)-\int\frac{d^{d+1}k'}{(2\pi)^{d+1}}
		V_{k,k'} \tau_3 {\hat{\cal G}}_s(k'+q) {\hat \Gamma}_s(k',k'+q) {\hat{\cal G}}_s(k') \tau_3
\end{equation} 
We shall use the notation $q\equiv ({\bf q},q_0\equiv \omega)$, where $q_0$ is now the Wick rotated frequency. The three Pauli matrices are the conventional $\tau_1$, $\tau_2$, and $\tau_3$. The equation is of course a matrix equation. Near a collective mode, ${\hat \Gamma}_s(k,k+q)$ goes to $\infty$.  The mode spectrum
is then determined from the equation
\begin{equation}
{\hat \Gamma}_s(k,k+q)=-\int\frac{d^{d+1}k'}{(2\pi)^{d+1}} V_{k,k'} 
		\tau_3 { \hat{\cal G}}_s(k'+q) {\hat \Gamma}_s(k',k'+q) {\hat{\cal G}}_s(k') \tau_3
\end{equation}

Consider S-wave pairing where $V_{k,k'}= -\lambda<0$ and therefore
${\hat \Gamma}_s(q) = {\hat \Gamma}_s(k,k+q)$.  Close to the phase mode, the vertex is
a superposition of $\tau_3$ and $\tau_2$, but the coefficient of $\tau_2$
term diverges.  Thus, the spectrum satisfies the following equation:
\begin{equation}
\tau_2 \approx \int\frac{d^{d+1}k'}{(2\pi)^{d+1}} \lambda 
		\tau_3 {\hat{\cal G}}_s(k'+q) \tau_2 {\hat{\cal G}}_s(k') \tau_3
\end{equation}
The right hand side of the above equation also contains a $\tau_3$ term,
but it is negligible in deciding the spectrum.  Substituting the expression
for the Green's function 
\begin{equation}
{\hat{\cal G}}_s^{-1}(k)=\left( \begin{array}{cc} {\cal G}^{-1}(k) & \Delta \\
		\Delta & -{\cal G}^{-1}(-k) \end{array} \right), 
\end{equation}
we get 
\begin{equation}
1 \approx \lambda
\int\frac{d^{d+1}k}{(2\pi)^{d+1}} 	\frac{{\bf \cal G}^{-1}(k+q)
{\bf \cal G}^{-1}(-k)+\Delta^2}
	{[|{\bf \cal G}(k+q)|^{-2}+\Delta^2][|{\bf \cal G}(k)|^{-2}+\Delta^2]}.
\end{equation}
Subtracting the gap equation
\begin{equation}
1=\lambda
\int\frac{d^{d+1}k}{(2\pi)^{d+1}} \frac{1}{|{\bf \cal G}(k)|^{-2}+\Delta^2},
\end{equation}
we get
\begin{equation}
0=\int\frac{d^{d+1}k}{(2\pi)^{d+1}} 
\frac{[{\bf \cal G}^{-1}(k+\frac{q}{2})-{\bf \cal G}^{-1}(k-\frac{q}{2})]
	[{\bf \cal G}^{-1}(-k-\frac{q}{2})-{\bf \cal G}^{-1}(-k+\frac{q}{2})]}
{[|{\bf \cal G}(k+\frac{q}{2})|^{-2}+\Delta^2][|{\bf \cal G}(k-\frac{q}{2})|^{-2}+\Delta^2]}.
\end{equation}
In obtaining this equation, we have assumed particle-hole symmetry and have
used the property ${ \cal G}(k)=-{ \cal G}(k')$ when $k_0=-k'_0$ and
$\varepsilon_{\bf k}=-\varepsilon_{\bf k'}$.

In our model, the normal state Green's function is given by
\begin{equation}
{\cal G}^{-1}(k)=g^{-1}(\alpha) \omega_c^\alpha e^{i\frac{\alpha\pi}{2}{\rm Sign}(k_0)}
	[ik_0-\varepsilon_{\bf k}]^{1-\alpha}.
\end{equation}
Note that the imaginary part of ${\cal G}^{-1}(k)$ changes sign at $k_0=0$.
For small $q$ ($q_0>0$), we use the following approximations
\begin{eqnarray*}
{\cal G}^{-1}(k+\frac{q}{2})-{\cal G}^{-1}(k-\frac{q}{2}) \approx
\left\{ \begin{array}{cc}
2i g(\alpha)^{-1}\sin(\frac{\alpha\pi}{2}) \omega_c^\alpha |\varepsilon_{\bf k}|^{1-\alpha} $ for $ |k_0|<\frac{q_0}{2}, \\
\frac{(1-\alpha) e^{i\frac{\alpha\pi}{2}} \omega_c^\alpha}
	{g(\alpha)(ik_0-\varepsilon_{\bf k})^\alpha} [iq_0-{\bf q} \cdot {\bf v}_F({\bf k})]
	$ for $ |k_0|>\frac{q_0}{2}
\end{array}  \right. 
\end{eqnarray*}
and linearize the spectrum close to the Fermi surface.  We get
\begin{eqnarray}
0&=&  4 q_0\sin^2\left(\frac{\alpha\pi}{2}\right)
\int \frac{	|\epsilon|^{2-2\alpha} d\epsilon} {
	[g^{-2}(\alpha)  \omega_c^{2\alpha}
|\epsilon|^{2-2\alpha}+\Delta^2]^2}\nonumber \\  
&&+ (q_0^2+\overline{|{\bf v}_F
\cdot {q}|^2}) (1-\alpha)^2
\int \frac{	 d\epsilon d\omega }{ |\epsilon^2+\omega^2|^\alpha
	[g^{-2}(\alpha) \omega_c^{2\alpha} |\epsilon^2+\omega^2|^{1-\alpha}
 	+\Delta^2]^2} .
\end{eqnarray}
The limits of the integrals can be set to $\pm \infty$ with impunity. The notation $\overline{A}$ implies average over the Fermi surface. Finally, the spectrum for the phase mode is given by
\begin{equation}
2 \alpha \sin\left(\frac{\alpha\pi}{2}\right) B\left(\frac{3-2\alpha}{2-2\alpha},\frac{1-2\alpha}{2-2\alpha}\right)
	 \left(\frac{g(\alpha)\Delta}{\omega_c}\right)^\frac{\alpha}{1-\alpha}
		q_0 \Delta +
(1-\alpha)^2 (q_0^2+\overline{|v_F \cdot q|^2})=0.
\end{equation}

When $\alpha \ll 1$, to leading order in $\alpha$,
after the analytic continuation $\Omega=iq_0$, this equation becomes
\begin{equation}
-i \frac{\alpha^2 \pi^2}{2} \Omega \Delta -\Omega^2+\overline{|{\bf v}_F \cdot {\bf q}|^2}=0
\end{equation}
There are two solutions 
\begin{equation}
4\Omega_{1,2}=-i \alpha^2 \pi^2 \Delta \pm \sqrt{16\overline{|{\bf v}_F \cdot {\bf q}|^2}-\alpha^4 \pi^4 \Delta^2}.
\end{equation}
A propagating mode can exist only if 
\begin{equation}
q\xi_0 \ge {\pi^2\alpha^2\over 4}{1\over \sqrt{\overline{({\hat v}_F\cdot{\hat q})^2}}},
\end{equation}
where $\xi_0 =v_F/\Delta$ is the conventional definition of the coherence length.
For a fixed $\alpha$, the mode is entirely overdamped as $q\to 0$. 
The damping of the collective phase mode is due to the coupling to the gapless
single particle excitations of the non-Fermi liquid. As we
have shown from calculations of the density of states, such gaplessness persists
even in the superconducting state. In the presence of long range Coulomb
interactions, charge fluctuations associated with the phase mode are
energetically costly and the phase mode is irrelevant for superconductivity.

%% file: anomaly6.tex
It has been emphasized\cite{Chakravarty1} that the Josephson pair-tunneling 
Hamiltonian (pair-tunneling action in the language of the present paper) used in the
interlayer tunneling of high temperature superconductors is not consistent with Fermi liquid
theory, but requires a non-Fermi liquid Green's function as discussed in the present paper.
Such a Hamiltonian can be presently understood only in the spirit of an effective field
theory; similar arguments were also briefly described in Ref.~\cite{Chakravarty2}. In
other words, we construct a low energy Hamiltonian based on certain general
principles, such as symmetry. This approach is necessary because we do not, as
yet, have a precise microscopic understanding of  non-Fermi liquid theory. Thus,
we are unable to {\sl explicitly} integrate out the high energy degrees freedom to
arrive at this Hamiltonian. The difficulty is also compounded by the fact that
interesting results arise only if,  at the same time, we can  suppress single
particle tunneling between layers for which we can only give  qualitative
reasons. In particular, the incoherence of single particle tunneling between the
layers, for sufficiently large $\alpha$, (See also the section on spin-charge
separation, Sec. VIII.) is the clue that such a term should not be included in
the effective low energy theory. This appears   natural to us because an
incoherent process is not a stationary process and its time evolution is not an
oscillating exponential. The essential reason as to why single particle tunneling
is incoherent is that the electronic quasipartcles are not well defined in a
non-Fermi liquid, and the amplitude has to be reconstructed for an electron to
tunnel between the layers. Such a process necessarily involves overlap of
$N$-particle wave functions. In contrast, the superconducting wave function is an
antisymmetrized wave function of Cooper pairs, and the tunneling matrix element
only involves the overlap of a pair. It is the nature of a Bose condensed state
that all pairs behave the same way. Thus, tunneling of Cooper pairs cannot be
plagued with orthogonality catastrophe of $N$-particle wave functions. The aim of
the present section is to give reasons as to why the pair tunneling Hamiltonian
is reasonable from the point of view of effective field theory.

It is essential to realize that conventional Josephson effect in a Fermi 
liquid cannot be recast in the form of an effective
Hamiltonian\cite{Chakravarty1}. The well known expression for the critical
current for two conventional superconductors with equal gaps\cite{Ambegaokar}, 
\begin{equation}
I_c=\frac{\pi \Delta (T)}{2eR_{\rm N}}\tanh ( {\Delta (T)\over 2T} ),
\end{equation}
where $R_{\rm N}$ is the normal state tunneling resistance, is unusual in two
respects. As $T\to 0$, $I_c$ is a nonanalytic function of the
gap ($\propto |\Delta|$), and is analytic only
close to $T_c$ (Note that, close to $T_c$, the magnitude of the Josephson
effect, derived from Ginzburg-Landau theory\cite{Aslamasov}, is proportional to
$\Delta (\Delta / T)$ and the analytic result is due to the thermal smearing.). 
It is also
remarkable that $I_c$ is independent of the cutoff,
$\omega_D$, the Debye energy. The dependence on the Fermi energy is buried in the
definition of  $R_{\rm N}$. This is a manifestation of Fermi liquid theory.  
Consider the
expression for the critical current  at $T=0$, given by 
\begin{equation}
I_c \propto \sum_{\bf k,q}|T_{\bf k,q}|^2{|\Delta_{\bf k}|\over E_{\bf k}} 
{|\Delta_{\bf q}|\over E_{\bf q}}{1\over E_{\bf k}+E_{\bf q}},
\end{equation}
where $E_{\bf k}=\sqrt{\varepsilon ({\bf k})^2+|\Delta_{\bf k}|^2}$. The 
integration over
the momenta can be converted to energy integrations which extend only to the Debye
energy. If the energy integrations can be extended to infinity with impunity, it
can be  seen by scaling the energy variables that $I_c\propto |\Delta|$; the
correction is of order $({\Delta\over \omega_D})$. This result  hinges
on the Fermi-liquid scaling of the single particle spectral function: 
$ A(\Lambda k,\Lambda \omega)= \Lambda^{-1}A(k,\omega)$. Any non
zero value of the scaling exponent, $\alpha$, will result in a cutoff dependence 
of the
critical current $I_c$, and if this cutoff is much larger than the gap, the
critical current can be an analytic function of the gap, proportional to
$\Delta (\Delta/\omega_c)$. This allows us to reformulate Josephson effect in
terms of an instantaneous pair tunneling hamiltonian on energy scales smaller
than the cutoff.

In fact, explicit calculations\cite{Chakravarty1} using the non-Fermi liquid 
spectral function of the present paper  leads to
\begin{equation}
I_c\propto \langle |T_{\bf k,q}|^2\rangle v_{\rm F}^{-2} \Delta ({\Delta\over
\omega_c}) f(\alpha, {\Delta\over \omega_c}), 
\end{equation}
where the angular brackets imply an energy average and 
$f(\alpha,{\Delta\over \omega_c})$ is a complicated 
function. When ${\Delta\over \omega_c} \ll 1$, 
$f(0,\Delta/\omega_c)=\omega_c/\Delta$, and we recover the Fermi liquid result.
For $\alpha >1/4$, the dependence of $f(\alpha,{\Delta\over \omega_c})$ 
on ${\Delta\over \omega_c}$ disappears entirely
in the limit ${\Delta\over \omega_c}\ll 1$, and it is a function of $\alpha$ alone. 
A closed form expression in the regime ${1\over 2}>\alpha>{1\over 4}$ was noted
earlier\cite{Chakravarty1}.  For $\alpha$ not too small, but less than ${1\over 4}$, the
dependence on ${\Delta\over \omega_c}$ is weak. 

We now construct a pair tunneling Hamiltonian. This Hamiltonian is certainly 
justified for $\alpha>{1\over 4}$, which is also the threshold at which
incoherence sets in in the spin-chrage separated model (See Sec. VIII); for the
case without spin-charge separation the threshold of incoherence is
$\alpha={1\over 2}$. In the tunneling between the layers, the transverse wave
vector of an electron is conserved, and this conservation is an exact symmetry. Thus, consider the
effective action
\begin{equation}
S_I^J=\int dt' dt d^{2}{\bf k}_1 d^{2}{\bf k}_2 
{\tilde g}({\bf k}_1, {\bf k}_2,|t-t'|)\left[\psi^{*(1)}_{{\bf
k}_1\sigma_1}(t)\psi_{{\bf k}_2\sigma_2}^{*(1)}(t')
\psi^{(2)}_{{\bf k}_2\sigma_2}(t)\psi^{(2)}_{{\bf k}_1\sigma_1}(t')+(1\to 2)\right].
\end{equation}
In view of the discussion above, we may set 
${\tilde g}({\bf k}_1, {\bf k}_2, |t-t'|)={g({\bf k}_1,{\bf k}_2)\over
\omega_c}\delta(t-t')$. Because in a non-Fermi lquid the tunneling process is so widely and uniformly spread over frequency  that the Kernel is instantaneous. If we now repeat the scaling arguments of Sec. IIE, we
see that the scaling dimension of this term at the non-Fermi liquid fixed point
is $s^{-(1-2\alpha)}$. This term is therefore {\sl relevant}, diverging as $s\to
0$, as long as
$\alpha < {1\over 2}$ and becomes marginal at $\alpha={1\over 2}$. In contrast,
we have seen in Sec. IIE that the conventional BCS pairing interaction (the
in-plane mechanism in the interlayer tunneling theory), which involves
integration over three independent wave vectors, is {\sl irrelevant}, scaling as
$s^{2\alpha}$, hence vanishing as $s\to 0$. Indeed, we have seen in Sec. III that
a coupling larger than a critical coupling is necessary to drive a superconducting
transition, if we use a BCS pairing kernel. 

It is reasonable to assume a separable form for $g({\bf k}_1,{\bf k}_2)$: 
\begin{equation}
g({\bf k}_1,{\bf k}_2)=-t_{\perp}({\bf k}_1)t_{\perp}({\bf k}_2)
\end{equation}
where $t_{\perp}({\bf k})$ is the single electron tunneling matrix element 
that can be determined from microscopic band structure calculations. An expression for
$t_{\perp}({\bf k})$  based on   general considerations of symmetry and
analyticity was proposed in Ref~\cite{Chakravarty2}; it is
\begin{equation}
t_{\perp}({\bf k})={t_{\perp}\over 4}\left(\cos k_xa - \cos k_ya\right)^2.
\end{equation}
This expression for $t_{\perp}({\bf k})$ is now confirmed from more detailed 
considerations of actual band structure computations\cite{Ole}. 

As in the original BCS theory, the pair tunneling action can be simplified, and 
a reduced action can be used for computations within the mean field
approximation. This reduced Hamiltonian, which involves only pairing of states
$({\bf k}\uparrow)$ with $({-\bf k}\downarrow)$, is 
\begin{equation}
-\int dt d^{2}{\bf k} 
{t_{\perp}^2({\bf k})\over \omega_c}
\left[\psi^{*(1)}_{{\bf k}\uparrow}(t)\psi_{-{\bf k}\downarrow}^{*(1)}(t)
\psi^{(2)}_{-{\bf k}\downarrow}(t)\psi^{(2)}_{{\bf k}\uparrow}(t')+(1\to 2)\right].
\end{equation}
Note that it is misleading to consider the relevancy of this reduced action 
with the dimension of the fermion operators determined from the fixed action of
the normal state. The reduced action is equal to the original one only in the BCS
subspace.

In addition to pair tunneling, we may also construct a particle-hole pair tunneling term
given by
\begin{equation}
S_I^{p-h}=\int dt' dt d^{2}{\bf k}_1 d^{2}{\bf k}_2 
{\tilde g}({\bf k}_1, {\bf k}_2,|t-t'|)\left[\psi^{*(1)}_{{\bf
k}_1\sigma_1}(t)\psi_{{\bf k}_1\sigma_1}^{(1)}(t')
\psi^{(2)}_{{\bf k}_2\sigma_2}(t)\psi^{*(2)}_{{\bf k}_2\sigma_2}(t')+(1\to 2)\right].
\end{equation}
Similar to the single particle tunneling, the importance of this term is difficult to decide
because this process is likely to be highly incoherent in a non-Fermi liquid for precisely
the same reasons. In other words, the time dependence of the kernel may not be replaced by
$\delta(t-t')$. If, on the other hand, the system has a tendency to form a spin density
wave, this term may be more important than the pair tunneling term.

%% file: anomaly7.tex
For interlayer tunneling in which there are both in-plane attractive
interactions and Josephson pair hopping between the planes, we take $S_0$ to be
\begin{equation}
S_0=-{1\over \beta}\sum_{k,\sigma,\omega_n,i}
{\cal
G}^{-1}({\bf k},\omega_n)\psi_{{\bf k}\sigma}^{(i)*}(\omega_n)
\psi_{{\bf k}\sigma}^{(i)}(\omega_n).
\end{equation}
Similarly,
\begin{equation}
S_I^{\rm i-p}=-{1\over
\beta^3}\sum_{{\bf k},{\bf k'},\omega,\omega_1,\omega_2}
V_{{\bf k},{\bf k'}}\left[\psi_{{\bf k}\uparrow}^{(1)*}(\omega_1)
\psi_{-{\bf k}\downarrow}^{(1)*}(\omega-\omega_1)
\psi_{-{\bf k'}\downarrow}^{(1)}(\omega-\omega_2)
\psi_{{\bf k'}\uparrow}^{(1)}(\omega_2)+(1\to 2)\right]
\end{equation}
and
\begin{equation}
S_I^{\rm J}=-{1\over
\beta^3}\sum_{{\bf k},\omega,\omega_1,\omega_2}
T_J({\bf k})\left[\psi_{{\bf k}\uparrow}^{(1)*}(\omega_1)
\psi_{-{\bf k}\downarrow}^{(1)*}(\omega-\omega_1)
\psi_{-{\bf k}\downarrow}^{(2)}(\omega-\omega_2)
\psi_{{\bf k}\uparrow}^{(2)}(\omega_2)+(1\to 2)\right]
\end{equation}
The term $S_I^{\rm i-p}$ corresponds to a BCS in-plane pairing kernel 
as discussed in Secs. III and IV, while the term $S_I^{\rm J}$ is the interlayer
pair tunneling term introduced in Sec. V. This model was discussed in some detail
in Ref.~\cite{Chakravarty2}, where, for simplicity, the single particle
spectrum was assumed to be that of a Fermi liquid. The assumption was that
for in-plane motions the quasiparticles are recovered in the
superconducting state. This assumption is not entirely justified. As we have seen
in the previous sections, the spectral function is in fact quite broad; see
Fig.~(6). Thus, in the present section we include properly the non-Fermi
liquid features.

The Hubbard-Stratonovich transformation for the combined action is not entirely
straightforward due to the local-in-{\bf k} pair hopping term. First, let us
rewrite the Josephson term as
\begin{equation}
S_I^{\rm J}=2\sum_{k,\omega}\left[\hat A({\bf k},\omega)^{*}\hat A({\bf
k},\omega)-\hat R^2({\bf k},\omega)+\hat I^2({\bf k},\omega)\right],
\end{equation}
where we have ignored an unimportant single particle term that can be
incorporated in the band structure. Here,
\begin{eqnarray}
\hat A({\bf k},\omega)^{*}&=&{1\over 2}\sqrt{T_J({\bf k})\over
\beta^3}\sum_{\omega_1}
\left[\psi_{{\bf k}\uparrow}^{(1)*}(\omega_1)\psi_{-{\bf
k}\downarrow}^{(1)*}(\omega-\omega_1)+(1\to 2)\right]\\
\hat R({\bf k},\omega)&=&{1\over 4}\sqrt{T_J({\bf k})\over
\beta^3}\sum_{\omega_1}
\left[\psi_{{\bf k}\uparrow}^{(1)*}(\omega_1)\psi_{-{\bf
k}\downarrow}^{(1)*}(\omega-\omega_1)+\psi_{{-\bf
k}\downarrow}^{(1)}(\omega-\omega_1)\psi_{{\bf k}\uparrow}^{(1)}(\omega_1)-(1\to 2)\right]\\
\hat I({\bf k},\omega)&=&{1\over 4}\sqrt{T_J({\bf k})\over
\beta^3}\sum_{\omega_1}
\left[\psi_{{\bf k}\uparrow}^{(1)*}(\omega_1)\psi_{-{\bf
k}\downarrow}^{(1)*}(\omega-\omega_1)-\psi_{{-\bf
k}\downarrow}^{(1)}(\omega-\omega_1)\psi_{{\bf k}\uparrow}^{(1)}(\omega_1)-(1\to 2)\right]\\
\end{eqnarray}
The in-plane terms are decoupled, as usual, by using
\begin{eqnarray}
\hat B_1^{*}(\omega)&=&{1\over \beta^{3/2}}\sum_{k,\omega}\lambda ({\bf k})
\psi_{{\bf k}\uparrow}^{(1)*}(\omega)\psi_{-{\bf
k}\downarrow}^{(1)*}(\omega-\omega_1)\\
\hat B_2^{*}(\omega)&=&{1\over \beta^{3/2}}\sum_{k,\omega}\lambda ({\bf k})
\psi_{{\bf k}\uparrow}^{(2)*}(\omega)\psi_{-{\bf
k}\downarrow}^{(2)*}(\omega-\omega_1)
\end{eqnarray}
We also include in the functional integral the factor
\begin{equation}
\prod_\omega\exp\left[-B_1^*(\omega)B_1(\omega)-B_2^*(\omega)B_2(\omega)\right],
\end{equation}
and the factor
\begin{equation}
\prod_{k,\omega}\exp\left[-2A^*({\bf k},\omega)A({\bf k},\omega)
-2R^2(k,\omega)-2I^2(k,\omega)\right],
\end{equation}
and integrate over $B_1(\omega)$, $B_1^{*}(\omega)$, $B_2(\omega)$,
$B_2^{*}(\omega)$,
$A({\bf k},\omega)$,
$A^{*}({\bf k},\omega)$, and over two real fields $R({\bf k},\omega)$, and
$I({\bf k},\omega)$. A set of linear shifts
\begin{eqnarray}
B_1(\omega)&\to&B_1(\omega)-\hat B_1(\omega)\\
B_2(\omega)&\to& B_2(\omega)-\hat B_2(\omega)\\
A_k&\to&A_k-\hat A({\bf k},\omega)\\
R_k&\to&R_k-i\hat R({\bf k},\omega)\\
I_k&\to&I_k-\hat I({\bf k},\omega)
\end{eqnarray}
eliminate all the quartic terms. The fermions can now be integrated out and the
action can be written as
\begin{eqnarray}
S=\sum_{\omega}\left[B_1^{*}(\omega)B_1(\omega)+B_2^{*}(\omega)B_2(\omega)\right]
&+&\sum_{k,\omega}\left[A^{*}({\bf k},\omega)A({\bf
k},\omega)+R^2({\bf k},\omega)+I^2({\bf k},\omega)\right]\nonumber
\\ &-&\ln{\rm Det}\left[M^{(1)}M^{(2)}\right],
\end{eqnarray}
where $M^{(1)}$ and $M^{(2)}$ are two $2\times 2$ matrices in {\bf k}-$\omega$
space. The problem can be simplified if we note that at the saddle point
\begin{eqnarray}
\lambda({\bf k})B_1(\omega)&+&\sqrt{T_J({\bf k})}\left(A({\bf k},\omega)+{iR({\bf
k},\omega)\over 2}+{I({\bf k},\omega)\over 2}\right)=\nonumber \\
&&\left[\lambda({\bf k})B_1^{*}(\omega)+\sqrt{T_J({\bf k})}\left(A^{*}({\bf
k},\omega)+{iR({\bf k},\omega)\over 2}-{I({\bf k},\omega)\over
2}\right)\right]^{*}.
\end{eqnarray}
Therefore $R({\bf k},\omega)=I({\bf k},\omega)=0$. At the saddle point the
determinant of the product of two matrices simplifies to
\begin{eqnarray}
{\rm Det}\left[M^{(1)}M^{(2)}\right]={1\over \beta^4}&&
\left[{\cal G}^{-1}({\bf k},\omega){\cal G}^{-1}(-{\bf k},\omega-\omega_1)+\beta^{-1}
|\lambda({\bf k})B_1+\sqrt{T_J({\bf
k})}A({\bf k})|^2\right]\nonumber \\
&\times&\left[{\cal G}^{-1}({\bf k},\omega){\cal G}^{-1}(-{\bf
k},\omega-\omega_1)+\beta^{-1} |\lambda({\bf k})B_2+\sqrt{T_J({\bf
k})}A({\bf k})|^2\right],
\end{eqnarray}
where all the fields are evaluated at $\omega=0$: $B_1\equiv B_1(0)$,
$B_2\equiv B_2(0)$, and $A({\bf k})\equiv A({\bf k},0)$. On taking functional
derivatives, we get,
\begin{eqnarray}
0={\delta S\over \delta B_1^{*}}=B_1-{1\over \beta}\sum_{k,\omega}
\frac{\lambda({\bf k})\left[\lambda({\bf k})B_1+\sqrt{T_J({\bf
k})}A({\bf k})\right]} {{\cal G}^{-1}({\bf k},\omega){\cal G}^{-1}(-{\bf
k},-\omega)+\beta^{-1} |\lambda({\bf k})B_1+\sqrt{T_J({\bf
k})}A({\bf k})|^2}\\
0={\delta S\over \delta A(k)^{*}}=2A(k)-{1\over \beta}\sum_{k,\omega}
\frac{\sqrt{T_J({\bf k})}\left[\lambda({\bf k})B_1+\sqrt{T_J({\bf k})}A({\bf
k})\right]} {{\cal G}^{-1}({\bf k},\omega){\cal G}^{-1}(-{\bf
k},-\omega)+\beta^{-1} |\lambda({\bf k})B_1+\sqrt{T_J({\bf
k})}A({\bf k})|^2} \\
0={\delta S\over \delta B_2^{*}}=B_2-{1\over \beta}\sum_{k,\omega}
\frac{\lambda({\bf k})\left[\lambda({\bf k})B_2+\sqrt{T_J({\bf k})}A({\bf
k})\right]} {{\cal G}^{-1}({\bf k},\omega){\cal G}^{-1}(-{\bf
k},-\omega)+\beta^{-1} |\lambda({\bf k})B_1+\sqrt{T_J({\bf
k})}A({\bf k})|^2}.
\end{eqnarray}
The saddle point is at $B_1=B_2\equiv B$. Now if we define the total gap
$\Delta_{\bf k}$ by
\begin{equation}
\Delta_{\bf k}=\lambda({\bf k})B+\sqrt{T_J({\bf k})}A({\bf k}),
\end{equation}
then
\begin{equation}
\Delta_{\bf k}\left[1-\chi({\bf k})
T_J({\bf k})\right]=\lambda({\bf k})\sum_{k'}\lambda({\bf k}')\Delta_{{\bf
k}'}\chi({\bf k}').
\end{equation}
The pair susceptibility in terms of the total gap is the same as defined
earlier. Note that the only difference from the result in
Ref.~\cite{Chakravarty2} is the presence of non-Fermi liquid effects in
the pair susceptibility. The derivation may appear to be a little more involved
than necessary, but the formalism that we have set up could be quite useful if
one is interested in the theory of collective modes\cite{Lan}. This equation can
be solved for any symmetry of the pairing interactions.

At $T=0$, this gap equation can be easily solved  on the Fermi
surface. Consider the
anisotropic $s$-wave case\cite{Chakravarty2}: $\lambda({\bf k})=\sqrt{\lambda}$.
At the Fermi surface
\begin{equation}
\Delta_{{\bf k}_F}=\Delta_0+T_J({\bf k}_F)\Delta_{{\bf k}_F}\chi({\bf k}_F),
\end{equation}
where 
\begin{equation}
\Delta_0=\lambda\sum_{k'}\Delta_{{\bf k}'}\chi({\bf k}')
\end{equation}
First consider the case $\alpha \ll 1/2$. Then, we can let the cutoff to
$\infty$, because the integral converges. We get, if the Josephson pair
hopping term dominates,
\begin{equation}
\Delta_{{\bf k}_F}\approx\omega_c g^{-1}(\alpha)\left[{T_J({\bf k}_F)\over
\omega_c}\right]^{1-\alpha\over 1-2\alpha}\left[{1\over 2(1-\alpha)\sin{\pi\over
2(1-\alpha)}}\right]^{1-\alpha\over 1-2\alpha}.
\end{equation}
This, as noted earlier\cite{Sudbo}, is encouraging, because local-in-{\bf k} interlayer
hopping produce a gap on the Fermi surface and there is no need for a critical
coupling, necessary for purely in-plane mechanism discussed earlier. As
$\alpha\to 1/2$, the above approximation of neglecting the cutoff $\omega_D$
cannot be used. However, the inegral can be easily approximated with the cutoff
in place. The result, for $\alpha\to 1/2$, is
\begin{equation}
\Delta_{{\bf k}_F}\approx\omega_c g^{-1}(\alpha)\left({\omega_D\over
\omega_c}\right)^{1-\alpha}e^{-{\pi\omega_c(1-\alpha)\over T_J({\bf k}_F)}}.
\end{equation}
Note that the precise results are not in agreement with those 
of Ref.~\cite{Sudbo}. This is due to an incorrect choice of the phase of the
Green's function.

The density of states as $\omega\to 0$ can also be estimated, and surprisingly
the behavior found earlier for the in-plane pairing is unchanged. We find for
$\alpha \ll 1/2$
and $\omega \to 0$:
\begin{equation}
{N(\omega)-N(0)\over \rho}= \left(\omega\over
x\right)^{2-2\alpha}\left({2-2\alpha\over 2-\alpha}\right),
\end{equation}
where  $\tilde{\Delta}_0 < x < \tilde{\Delta}_{\rm max}$; $\tilde{\Delta}_{\rm max}=
\max[\Delta_{\rm eff}({\bf k})]$, and $\tilde{\Delta}_0=\min[\Delta_{\rm eff}({\bf k})]$. Thus,
there is a finite density of states at $\omega=0$, and the low frequency
behavior is $\omega^{2-\alpha}$.

It is interesting to compute  tunneling density of states with a realistic 
Fermi surface corresponding to a two-layer high temperature superconductor, such
as Bi2212 or YBCO, assuming an $s$-wave in-plane pairing kernel. For this we
choose\cite{band}
\begin{equation}
\varepsilon_k=-0.5(\cos k_xa + \cos k_ya) + 0.45 \cos k_xa \cos k_ya,
\end{equation}
where $a$ is the lattice constant. The pair-hopping term $T_J({\bf k})$ was chosen to be\cite{Chakravarty2} 
\begin{equation}
T_J({\bf k})={T_J\over 16}\left(\cos k_xa - \cos k_ya\right)^4.
\end{equation}
The other parameters are as follows: $T_J=0.05$ eV, $\varepsilon_F=-0.315$ eV, $\Delta_0=0.01$ eV, and $\omega_c=0.15$ eV. The results are shown in Fig. (9). The tunneling density of states are qualitatively similar to those of Renner and Fischer\cite{Renner}.
\begin{figure}[htb]
\centerline{\epsfxsize 9cm
\epsffile{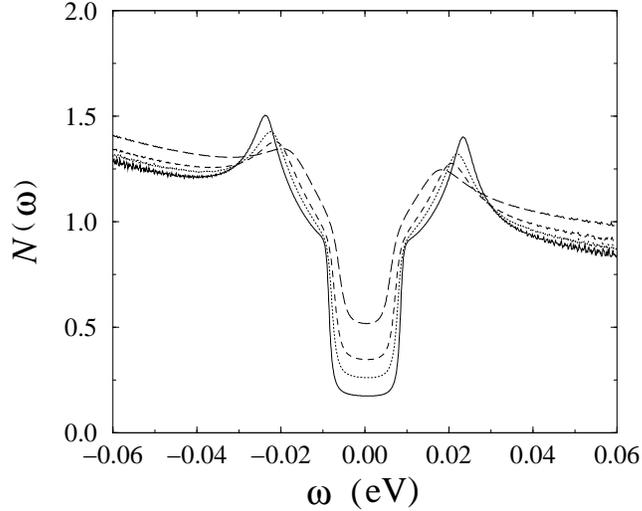}
}
\caption{Superconducting density of states for interlayer tunneling with realistic band parameters.  The solid, dotted, dashed, long-dashed curves are for $\alpha=0.05, 0.075, 0.01, 0.15$.}
\label{f:9}
\end{figure}

%% file: anomaly8.tex
In this section we shall consider a simple model for spin-charge
separation\cite{Anderson3}. The time-ordered Green's function at $T=0$ is
\begin{equation}
G(k,\omega)=\frac{g(\alpha) e^{-i \frac{\pi\alpha}{2} {\rm Sign}(\omega)}}
	{\omega_c^\alpha (\omega-v_s k+i\delta \omega)^\frac{1}{2}
	 (\omega-v_\rho k+i\delta \omega)^{\frac{1}{2}-\alpha}},
\end{equation}
where $v_s$ and $v_\rho$ are the spin and charge velocities ($v_\rho>v_s$),
$k$ is measured from the Fermi surface, $g(\alpha)$ is the same
as before, and $\omega_c$ is the cutoff.
The corresponding spectral weight function is 
\begin{eqnarray}
A(k,\omega) &=&{1\over \pi}|{\rm Im} G(k,\omega)| \\
	&=&  \begin{array}{cc} 
 {1\over \pi}|G(k,\omega)| \sin(\frac{\pi\alpha}{2}),$ when $(\omega-v_\rho k)(\omega-v_s
k)>0
\\
 {1\over \pi}|G(k,\omega)|\cos(\frac{\pi\alpha}{2}),$ when $(\omega-v_\rho k)(\omega-v_s
k)<0
	\end{array} 
\end{eqnarray}
It has two power-law singularities and high frequency tails.  The spectral
weight in the region between the two singularities is larger 
by a factor of $\cot(\frac{\pi\alpha}{2})$.

The momentum distribution function can be obtained by the formula
$n(k)=\int^0_{-\omega_c} A(k,\omega) d\omega$ and yields the value $1/2$ for
$n(0)$.  At finite $k$, and  for $v_\rho|k|<<\omega_c$, we get
\begin{equation}
n(k) \approx \frac{1}{2} -\frac{{\rm Sign}(k)}{2} \left[ \left(\frac{1}{2}-\alpha\right)
	B\left(\frac{1}{2},1-\alpha\right) \left| \frac{(v_\rho-v_s)k}{\omega_c} \right|^\alpha
	-f\left(\alpha,\frac{v_\rho}{v_s}\right) 
	\left| \frac{v_s k}{\omega_c} \right|^\alpha\right],
\end{equation}
where the function $f(\alpha,y)$ is given by the integral
\begin{equation}
f(\alpha,y)=\alpha\int^1_0\frac{dx}
{x^\frac{1}{2}(x+y-1)^{\frac{1}{2}-\alpha}}.
\end{equation}
We recover the earlier result for $n(k)$ without spin-charge separation when we set 
$v_\rho=v_s$, which results in $f(\alpha,1)=1$.
The density of states, $n(\omega)$, related to the spectral function by the formula
$n(\omega)=\sum_k A(k,\omega)$,  
is constant when $v_{\rho}$ and $v_s$ are independent of $\omega$ and $k$.

If we consider, once again, the problem of tunneling between two planes, we find that
the inverse of the matrix Green's function, ${\hat G}_t(k,\omega)$, is
\begin{equation}
[{\hat G}_t(k,\omega)]^{-1}=\left(\begin{array}{cc} G^{-1}(k,\omega) &  -t_\perp(k) \\
			 -t_\perp(k) & G^{-1}(k,\omega) \end{array} \right).
\end{equation}
The quasiparticle poles are given by the equation $\det|G_t^{-1}(k,\omega)|=0$,
more explicitly by
\begin{equation}
\omega_c^{2\alpha} g^{-2}(\alpha) e^{i\alpha \pi {\rm Sign}(\omega)} 
(\omega-v_s k+i\omega\delta)(\omega-v_\rho k +i\omega\delta)^{1-2\alpha}=t^2_\perp(k).  
\end{equation}
From the phase and the amplitude of this equation, we get
\begin{eqnarray}
{\rm Sign}(\omega) \alpha \pi+\theta_s+\theta_\rho(1-2\alpha)&=&2N\pi, \\
\omega_c^{2\alpha} |\omega-v_s k| |\omega-v_\rho k|^{1-2\alpha}&=&g^2(\alpha)
t_\perp^2(k).
\end{eqnarray}
For $k>0$ the equation for the phase restricts the poles in the shaded
area shown in Fig(10). 
\begin{figure}[htb]
\centerline{\epsfxsize 9cm
\epsffile{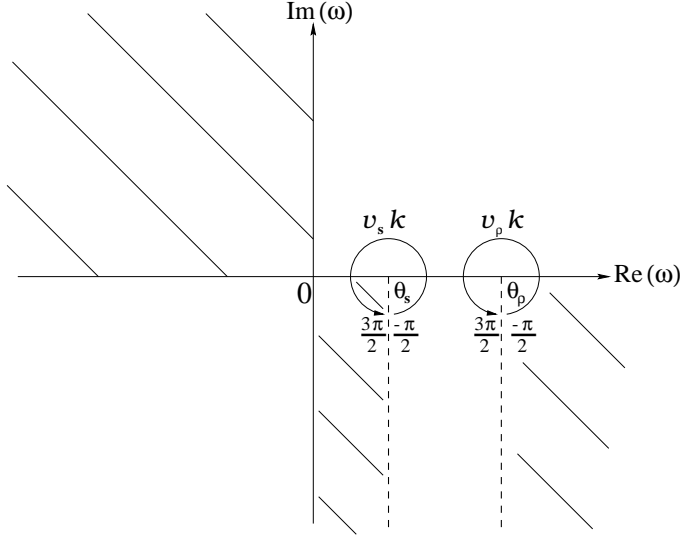}
}
\caption{Complex plane depiction of the singularities of the Green's function for the spin-charge separated model.}
\label{f:10}
\end{figure}
In general, there are two poles $\omega_{k}^+$ and $\omega_{k}^-$,
where ${\rm Re}\ \omega_{k}^+>v_\rho k$, ${\rm Im}\ \omega_{k}^+<0$, and
${\rm Re}\ \omega_{k}^-<v_s k$,
$\omega_{k}^- {\rm Im}\ \omega_{k}^-<0$ . However, when $\alpha>\frac{1}{4}$ and
$t_\perp(k)$ is less than a critical value, the pole $\omega_{k+}$ is absent, which
clearly indicates supression of coherent tunneling.  This is an important respect in
which the spin-charge separated case differs from the previously discussed case of no
spin-charge separation. To understand why the pole disappears, consider the equation for
the phase:
\begin{equation}
\alpha \pi+\theta_s+\theta_\rho(1-2\alpha)=0.
\end{equation}
Because $\theta_\rho>-\frac{\pi}{2}$, $\theta_s$ is bounded by $\theta_s \leq
\frac{\pi}{2}(1-4 \alpha)$.
When $\alpha>\frac{1}{4}$, this inequality confines the pole in
the shaded area as shown in Fig(11).
\begin{figure} 
\centerline{\epsfxsize 9cm
\epsffile{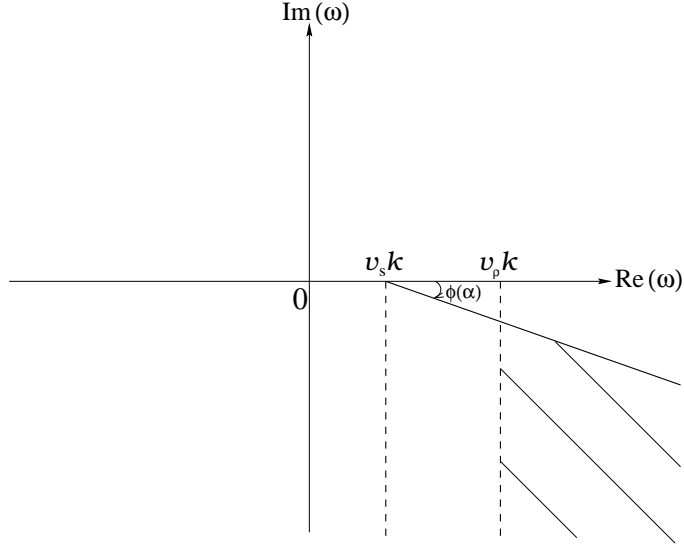}
}
\caption{The restriction of the pole in the shaded region of the complex plane.}
\label{f:11}
\end{figure}
For  $\omega$
belonging to the shaded area, the  left hand side of the amplitude
equation has the minimum:
$\omega_c^{2\alpha}(v_\rho k-v_s k)^{2(1-\alpha)}
|\sec(\phi(\alpha))| |\tan(\phi(\alpha))|^{1-2\alpha}$,
where $\phi(\alpha) = \frac{\pi}{2}(1-4 \alpha)$. 
For $g^{2}(\alpha)t^2_\perp(k)$  less than this minimum value, the equation for the amplitude can not
be satisfied.  So the pole $\omega_{k+}$ does not exist for $\alpha>\frac{1}{4}$
and $t_\perp(k)$ less than a critical value given by the minimum.

We now discuss the superconducting properties. The calculations are similar to those
discussed earlier. For S-wave pairing, the gap is 
\begin{equation}
\Delta=g^{-1}(\alpha)\omega_D\left[ \frac{h(\alpha,\frac{v_s}{v_\rho})}
	{h'(\alpha,\frac{v_s}{v_\rho}) 
	B(\frac{1}{1-\alpha},\frac{1-2\alpha}{1-\alpha})} 
\right]^{\frac{1-\alpha}{2\alpha}}
\left(1-\frac{\lambda_c}{\lambda}\right)^{\frac{1-\alpha}{2\alpha}},
\end{equation}
where
\begin{eqnarray*}
\lambda_c = \frac{\alpha \pi}
	{\rho g^2(\alpha) h(\alpha,\frac{v_s}{v_\rho})}
	\left(\frac{\omega_c}{\omega_D}\right)^{2\alpha},\\
h(\alpha,\beta) = \int_0^\frac{\pi}{2} \frac{\sec^{2\alpha}\theta d\theta}
	{\sqrt{\cos^2\theta+\beta^2\sin^2\theta}},\\
h'(\alpha,\beta) = \int_0^\frac{\pi}{2} \frac{d\theta}
	{[\cos^2\theta+\beta^2\sin^2\theta]^{\frac{1}{2-2\alpha}}}
\end{eqnarray*}
For $d$-wave, the gap is given by $\Delta_k=\Delta_0\cos\theta_k$, where the
amplitude $\Delta_0$ is
\begin{equation}
\Delta_0=g^{-1}(\alpha)\omega_D
\left[\frac{2h(\alpha,\frac{v_s}{v_\rho})}
		{h'(\alpha,\frac{v_s}{v_\rho})
 (1+\alpha)
	B(\frac{1}{1-\alpha},\frac{3}{2}-\frac{1}{1-\alpha})}
\right]^{\frac{1-\alpha}{2\alpha}}
\left(1-\frac{\lambda'_c}{\lambda}\right)^{\frac{1-\alpha}{2\alpha}}.
\end{equation}
The critical coupling  $\lambda'_c=2 \lambda_c$.
The equation for the transition temperature, $T_c$, is unchanged from the case without
spin-charge separation, provided the present definitions of the critical couplings,
$\lambda_c$ and $\lambda'_{c}$ are noted.

We find that the quasiparticles are damped, and the real part of their
energy is greater than $|v_\rho k|$.  Close to $\omega=0$, the density of
states, $N(\omega)$, is smaller than that of the normal state.  For $s$-wave
pairing, we get
\begin{equation}
N(0) = \rho \left[1-(1-\alpha)
		B\left(\frac{2-\alpha}{2-2\alpha},\frac{2-3\alpha}{2-2\alpha}\right)
		\left(\frac{g(\alpha)\Delta}{\omega_c}\sqrt{\frac{v_\rho}{v_s}}\right)
		^\frac{\alpha}{1-\alpha}\right]
\end{equation}
and
\begin{equation}
{N(\omega)-N(0)\over \rho} \propto \alpha
\frac{|\omega|^{2-2\alpha}\omega_c^{\alpha}} {\Delta^2}.
\end{equation}
For $d$-wave pairing, we get
\begin{equation}
N(0) = \rho \left[1-\frac{1}{2}
		B\left(\frac{1}{2-2\alpha},\frac{2-3\alpha}{2-2\alpha}\right)
		\left(\frac{g(\alpha)\Delta_0}{\omega_c}\sqrt{\frac{v_\rho}{v_s}}\right)
		^\frac{\alpha}{1-\alpha}\right]
\end{equation}
and
\begin{equation}
{N(\omega)-N(0)\over \rho} \propto \frac{|\omega|}{\Delta_0}.
\end{equation}
As before, the linearity of the $d$-wave density of
states is unaffected by $\alpha$.

We conclude that the  spin-charge separated model behaves qualitatively 
similarly to the
model without spin-charge separation. However, we found one important difference between
the two models with respect to the coherence of tunneling between two layers in the
normal state. In the spin-charge separated model, the coherence is absent if
$\alpha > {1\over 4}$, and the tunneling matrix element $t_\perp(k)$ is less than
a crtical value. Given the anisotropy of $t_\perp(k)$\cite{Chakravarty2}, this
can lead to interesting results.

%% file: anomaly9.tex
In the present section we shall summarize our work and point out the directions
in which our theory must be improved and extended. We have shown that the notion
of spectral anomaly is a powerful one. It allows us to formulate a framework
with which we can understand non-Fermi liquid effects on
superconductivity, regardless of the microscopic origin of the spectral
anomaly. We believe that any low energy effective field theory of gapless fermions
can be characterized by an anomalous spectral function that reflects the
critical state of the fermions. The results obtained from the spin-charge
separated model are qualitatively similar to those obtained from the model
wthout spin-charge separation. However, the single particle tunneling between
the layers shows a greater degree of incoherence in the spin-charge separated
model. We have also shown that the concept of spectral anomaly  provides a basis
for understanding the  features of the interlayer tunneling theory of high
temperature superconductors.

Spectral anomaly has two important effects. It makes the the single particle
spectrum incoherent while retaining some of the features of a Fermi liquid. The
spectrum is gapless and therefore the system responds as a metal: an infinitesimal
electric field generates a conducting response. We have also seen that one can
still define a surface in the momentum space at which the nature of the
excitations changes, as at the Fermi surface of a Fermi liquid. However,
the momentum distribution is not discontinuous at this surface but its
derivative is singular. Spectral anomaly  has also the negative effect that
conventional BCS pairing is irrelevant (in the renormalization group sense). We
have seen that one needs a coupling larger than a critical coupling to achieve
superconductivity. Nonetheless, the interlayer pair-tunneling Hamiltonian is
relevant (not marginal). This is particularly
interesting because many of the features are quite distinctive and reflect the
dominating effect of the interlayer tunneling rather than the small residual
in-plane attractive mechanism, which can of course have any symmetry. 
According to the present analysis, the c-axis tunneling density of states is
finite at zero frequency, but a well-defined peak still persists at a finite
frequency unless the anomaly exponent, $\alpha$, is very large. While such
effects are reported, it is not clear if these are experimental artifacts or not.
However, if such effects are not artifacts and can be measured precisely, we
can infer the value of $\alpha$ from such experiments.

In the present paper we have been unable to discuss the renormalization of the
composite operators such as current. Although we have used the Ward identity to
indicate how the vertex  should scale, it has not been possible to go
further. In our opinion, it is not possible to do
so without more powerful methods, such as the unitarity expansion formulated
many years ago by Gribov, Migdal, and Polyakov\cite{Gribov}. The renormalization
of composite operators is important because most experimental probes
couple to composite operators. In this respect, it is necessary to go beyond the
leading scaling forms, as the dependences on temperature and frequency are often
determined by the  the presence of
dangerously irrelevant operators as argued by Nayak and Wilczek\cite{Nayak}.

Our justification of the interlayer tunneling Hamiltonian consisted of a 
two step process. We first showed that the single
particle tunneling is incoherent in a non-Fermi liquid for sufficiently large
$\alpha$, which need not, however, be much larger than $1\over 2$ to $1\over 4$.
We then argued that because of {\sl incoherence} such a term should not be
included in the effective low energy theory. Renormalization group arguments,
similar to those discussed in our paper show, however, that such a term {\sl is}
relevant as long as $\alpha < 1$. Josephson pair tunneling was then argued not
to be severely affected by the spectral anomaly and can be cast into an effective
low energy Hamiltonian. It can be argued that it should be possible to begin
with a single particle tunneling Hamiltonian and arrive at a low energy theory
in which single particle tunneling is absent but pair tunneling is present as
a result of correlation effects that give rise to spectral anomaly. Our inability
to do so is intimately intertwined with the neglect  of vertex corrections in the tunneling process as was remarked by Wen\cite{Wen}.

Ultimately, it will be necessary to provide a microscopic derivation of the spectral anomaly. Beginning with the pioneering work of Anderson\cite{Anderson2} many authors have investigated the notion of tomographic Luttinger liquid in two dimensions. Similarly, many authors have focused on an underlying gauge theory of strongly correlated electrons that could possibly give rise to
non-Fermi liquid behavior. Recently Khlebnikov\cite{Khlebnikov} has approached interlayer pair tunneling from this perspective. Such ideas seem promising to us.

%% file: appendix.tex
\section{Spectral function I}
We derive here the spectral function corresponding to one particle green's function of
the interlayer tunneling problem discussed in Sec.~(\ref{sec-interlayer}).
\begin{equation}
A_{ij}(k,\omega)=-{1\over \pi} {\rm Sign}(\omega) {\rm
Im}\left[G(k,\omega)\right]_{ij},
\end{equation}
where for both the regions $\varepsilon(k) < \omega < 0$ and $\omega <
\varepsilon (k)$, $\omega < 0$,
\begin{equation}
{\rm Im} \left[G(k,\omega)\right]_{11}=\frac{\sin\left({\alpha\pi\over 2}\right)
D^-(k,\omega)^{1-\alpha}\left[t_{\perp}^2(k)+
D^-(k,\omega)^{2(1-\alpha)}\right]}
{\left[
D^-(k,\omega)^{2(1-\alpha)}-\cos(\alpha\pi)t_{\perp}^2(k)
\right]^2+\sin^2(\alpha\pi)
t_{\perp}^4(k)}.\label{eq:ImG}
\end{equation}
For the regions $\omega > 0$, $\omega > \varepsilon(k)$ and $0 < \omega <
\varepsilon(k)$, ${\rm Im} \left[G(k,\omega)\right]_{11}$ is negative of the
expression given in Eq. (~\ref{eq:ImG}).  In the more compact form, the spectral
function is
\begin{equation}
A_{11}(k,\omega)={1\over t_{\perp}(k)}F_{11}\left(\frac{\omega-\varepsilon(k)}{t_{\rm eff}(k)}\right),
\end{equation}
where the dimensionless function $F_{11}(x)$ is given by
\begin{equation}
F_{11}(x)=\frac{\sin\left({\alpha\pi\over 2}\right)|x|^{1-\alpha}[1+|x|^{2(1-\alpha)}]}
	{\pi[|x|^{4(1-\alpha)}-2\cos(\alpha\pi)|x|^{2(1-\alpha)}+1]}.
\end{equation}

 Similarly, the spectral function related
to tunneling between the layers is related to the off-diagonal element by
\begin{equation}
A_{12}(k,\omega)=-{1\over \pi} {\rm Sign}(\omega) {\rm
Im}\left[G(k,\omega)\right]_{12},
\end{equation}
where, for $\varepsilon(k) < \omega < 0$,
\begin{equation}
{\rm Im} \left[G(k,\omega)\right]_{12}=\frac{\sin(\alpha\pi)
D^-(k,\omega)^{2(1-\alpha)}t_{\perp}(k)}
{\left[
D^-(k,\omega)^{2(1-\alpha)}-\cos(\alpha\pi)t_{\perp}^2(k)
\right]^2+\sin^2(\alpha\pi)
t_{\perp}^4(k)}.\label{eq:ImG12}
\end{equation}
In the region $\omega < 0$, $\omega < \varepsilon (k)$, ${\rm Im}
\left[G(k,\omega)\right]_{12}$ is negative of that given in Eq.
(~\ref{eq:ImG12}). In the region $\omega > 0$, $\omega > \varepsilon (k)$,
${\rm Im} \left[G(k,\omega)\right]_{12}$ is negative of that given in 
Eq. (~\ref{eq:ImG12}).  Finally, in the region $\varepsilon (k) > \omega > 0$,
${\rm Im} \left[G(k,\omega)\right]_{12}$ is the same as that given in 
Eq. (~\ref{eq:ImG12}).

Therefore we have
\begin{equation}
A_{12}(k,\omega)={1\over t_{\perp}(k)}F_{12}\left(\frac{\omega-\varepsilon(k)}{t_{\rm eff}(k)}\right),
\end{equation}
where
\begin{equation}
F_{12}(x)=\frac{{\rm Sign}(x)\sin(\alpha\pi) |x|^{2(1-\alpha)}}
	{\pi\left[[|x|^{2(1-\alpha)}-\cos(\alpha\pi)]^2+1\right]}.
\end{equation}

\section{Spectral function II}
It is interesting to calculate the superconducting density of states and the
spectral function. First, consider the spectral function. 
The imaginary part of the matrix element,
$\left[G_s^{\rm R}(k,\omega)\right]_{11}$, of the retarded Green's function is
related to the spectral function by
\begin{equation}
A_s(k,\omega)=-{1\over \pi}{\rm Im} \left[G_s^{\rm R}(k,\omega)\right]_{11}.
\end{equation}
In terms of the time-ordered Green's function,
\begin{equation}
A_s(k,\omega)=-{1\over \pi}{\rm Sign (\omega)}{\rm Im} \left[G_s
(k,\omega)\right]_{11}.
\end{equation}
First, consider first $\omega < 0$. There are three separate case: (1)
$\varepsilon(k) < \omega$, (2) $\omega < \varepsilon(k) < -\omega$, and (3)
$-\omega <
\varepsilon(k)$.
\begin{enumerate}
\item $\varepsilon(k) < \omega$
\\
In this region, we have $\varepsilon(k) < \omega < 0 < -\omega$. Therefore,
both $\omega-\varepsilon(k) > 0$ and $ -\omega-\varepsilon(k) > 0$. Then,
\begin{equation}
{\rm Im} \left[G_s(k,\omega)\right]_{11}={\sin\left({\alpha\pi\over 2}\right)
D^+(k,\omega)^{1-\alpha}\over
[D^+(k,\omega)D^-(k,\omega)]^{(1-\alpha)}+\Delta_k^2}
\end{equation}
\item $\omega < \varepsilon(k) < -\omega$
\\
In this region, there are two possibilities, either $\omega < 0 <
\varepsilon(k) < -\omega $ or $\omega < \varepsilon(k) < 0  < -\omega $. If we
remember that we go counterclockwise around one branch point and clockwise
around the other, we get, for both of these cases,
\begin{equation}
{\rm Im} \left[G_s(k,\omega)\right]_{11}=\frac{\sin\left({\alpha\pi\over 2}\right)
D^+(k,\omega)^{1-\alpha}\left[\Delta_k^2+[D^+(k,\omega)D^-(k,\omega)]^{(1-\alpha)}\right]}
{\left[[D^+(k,\omega)D^-(k,\omega)]^{(1-\alpha)}-\cos(\alpha\pi)\Delta_k^2\right]^2
+\sin^2(\alpha\pi)\Delta_k^4}.
\end{equation}
\item $-\omega < \varepsilon (k)$
\\
In this region, both $\omega -\varepsilon (k) < 0$ and  $-\omega -\varepsilon (k)
< 0$. Therefore 
\begin{equation}
{\rm Im} \left[G_s(k,\omega)\right]_{11}={\sin\left({\alpha\pi\over 2}\right)
D^+(k,\omega)^{1-\alpha}\over
[D^+(k,\omega)D^-(k,\omega)]^{(1-\alpha)}+\Delta_k^2}
\end{equation}
\end{enumerate}

The case $\omega > 0$ can be treated identically and we give the results below,
written in a more concise form.
\begin{enumerate}
\item $|\varepsilon(k)| >\omega$.
\\
\begin{equation}
{\rm Im} \left[G_s(k,\omega)\right]_{11}=-{\sin\left({\alpha\pi\over 2}\right)
D^+(k,\omega)^{1-\alpha}\over
[D^+(k,\omega)D^-(k,\omega)]^{(1-\alpha)}+\Delta_k^2}
\end{equation}
\item $|\varepsilon(k)| < \omega$.
\\
\begin{equation}
{\rm Im} \left[G_s(k,\omega)\right]_{11}=-\frac{\sin\left({\alpha\pi\over 2}\right)
D^+(k,\omega)^{1-\alpha}\left[\Delta_k^2+[D^+(k,\omega)D^-(k,\omega)]^{(1-\alpha)}\right]}
{\left[[D^+(k,\omega)D^-(k,\omega)]^{(1-\alpha)}-\cos(\alpha\pi)\Delta_k^2\right]^2
+\sin^2(\alpha\pi)\Delta_k^4}.
\end{equation}
\end{enumerate}

In the scaled form, the spectral weight can be written as
\begin{equation}
A_s(k,\omega)={1\over |\Delta_k|}F_s\left(\frac{\omega}{|\Delta_{\rm eff}(k)|},\frac{\varepsilon(k)}
				{|\Delta_{\rm eff}(k)|}\right),
\end{equation}
where
\begin{equation}
F_s(x,y)=\frac{\sin\left({\alpha\pi\over 2}\right)}{\pi}\left[
\frac{\theta(|y|-|x|)|x+y|^{1-\alpha}}{|x^2-y^2|^{1-\alpha}+1}+
\frac{\theta(|x|-|y|)|x+y|^{1-\alpha}(1+|x^2-y^2|^{1-\alpha})}
{[|x^2-y^2|^{1-\alpha}-\cos(\alpha\pi)]^2+\sin^2(\alpha\pi)}\right].
\end{equation}

The BCS result, for comparison, is
\begin{equation}
A_{\rm BCS}(k,\omega)=u_k^2 \delta(\omega-E_k)+v_k^2\delta(\omega+E_k),
\end{equation}
where $E_k=\sqrt{\varepsilon^2(k)+\Delta_k^2}$, and
\begin{eqnarray}
u_k^2&=&{1\over 2}\left(1+{\varepsilon(k)\over E_k}\right), \\
v_k^2&=&{1\over 2}\left(1-{\varepsilon(k)\over E_k}\right).
\end{eqnarray}
The superconducting density of states, $N(\omega)$, is
\begin{equation}
N(\omega)=-{1\over \pi}{\rm Sign (\omega)}\sum_k{\rm Im}
\left[G_s (k,\omega)\right]_{11}.
\end{equation}